\pdfoutput=1

\documentclass[iop]{emulateapj}

\usepackage{natbib,epsfig,longtable,siunitx,url,graphicx,amsmath,footnote,float,xcolor}
\usepackage[colorlinks=true,linkcolor=blue,citecolor=blue]{hyperref}

\begin{document}

\title{Prevalence of Chaos in Planetary Systems Formed Through Embryo Accretion}

\author{Matthew S. Clement\altaffilmark{1} \& Nathan A. Kaib\altaffilmark{1}}

\altaffiltext{1}{HL Dodge Department of Physics Astronomy, University of Oklahoma, Norman, OK 73019, USA \& corresponding author email: matt.clement@ou.edu}

\begin{abstract}

The formation of the solar system's terrestrial planets has been numerically modeled in various works, and many other studies have been devoted to characterizing our modern planets' chaotic dynamical state. However, it is still not known whether our planets’ fragile chaotic state is an expected outcome of terrestrial planet accretion.   We use a suite of numerical simulations to present a detailed analysis and characterization of the dynamical chaos in 145 different systems produced via terrestrial planet formation in \citet{kaibcowan15}.  These systems were created in the presence of a fully formed Jupiter and Saturn, using a variety of different initial conditions.  They are not meant to provide a detailed replication of the actual present solar system, but rather serve as a sample of similar systems for comparison and analysis.  We find that dynamical chaos is prevalent in roughly half of the systems we form.  We show that this chaos disappears in the majority of such systems when Jupiter is removed, implying that the largest source of chaos is perturbations from Jupiter.  Chaos is most prevalent in systems that form 4 or 5 terrestrial planets.  Additionally, an eccentric Jupiter and Saturn is shown to enhance the prevalence of chaos in systems.  Furthermore, systems in our sample with a center of mass highly concentrated between $\sim$0.8--1.2 AU generally prove to be less chaotic than systems with more exotic mass distributions.  Through the process of evolving systems to the current epoch, we show that late instabilities are quite common in our systems.  Of greatest interest, many of the sources of chaos observed in our own solar system (such as the secularly driven chaos between Mercury and Jupiter) are shown to be common outcomes of terrestrial planetary formation.  Thus, consistent with previous studies such as \citet{laskar96}, the solar system's marginally stable, chaotic state may naturally arise from the process of terrestrial planet formation.
\break
\break
{\bf Keywords:} Chaos, Planetary Formation, Terrestrial Planets \hfill Received; Accepted

\end{abstract}

\section{Introduction}

Our four terrestrial planets are in a curious state where they are evolving chaotically, and are only marginally stable over time \citep{laskar96, laskar08, laskar09}.  This chaos is largely driven by interactions with the 4 giant planets.  However our understanding of the dynamical evolution of the gas giants, particularly Jupiter and Saturn, has changed drastically since the introduction of the Nice Model \citep{gomes05, mor05, tsi05}.

The classical model of terrestrial planetary formation, where planets form from a large number of small embryos and planetesimals that interact and slowly accrete, is the basis for numerous studies of planetary evolution \citep[e.g.][]{chambers01, obrien06, chambers07, ray09, kaibcowan15}.  Using direct observations of proto-stellar disks \citep{currie09}, it is clear that free gas disappears long before the epoch when Earth's isotope record indicates the conclusion of terrestrial planetary formation \citep{halliday08}.  For these reasons, a common initial condition taken when numerically forming the inner planets is a fully formed system of gas giants at their current orbital locations.  Many numerical models have produced planets using this method.  However, none to date have analyzed the chaotic nature of fully evolved accreted terrestrial planets up to the solar system's current epoch.  It should be noted that other works have modeled the outcome of terrestrial planetary formation up to 4.5 Gyr.  \citet{laskar00} evolved 5000 such systems from  10000 planetesimals and showed correlations between the resulting power-law orbital spacing and the initial mass distribution.  Furthermore, many works have performed integrations of the current solar system, finding solutions that showed both chaos and a very real possibility of future instabilities \citep{laskar08, laskar09}.  Our work is unique in that we take systems formed via direct numerical integration of planetary accretion, evolve them to the solar system's age, probe for chaos and its source, and draw parallels to the actual solar system.

Although the classical terrestrial planet formation model has succeeded in replicating many of the inner solar system’s features, the mass of Mars remains largely unexplained \citep{chambers01, obrien06, chambers07, ray09, kaibcowan15}.  Known as the Mars mass deficit problem, most simulations routinely produce Mars analogues which are too massive by about an order of magnitude. \citet{tach} argue for an early inward, and subsequent outward migration of a fully formed Jupiter, which results in a truncation of the proto-planetary disc at 1 AU prior to terrestrial planetary formation.  If correct, this ``Grand Tack Model'' would explain the peculiar mass distribution observed in our inner solar system.  Another interesting solution involves local depletion of the disc in the vicinity of Mars's orbit \citep{iz14}.  A detailed investigation of the Mars mass deficit problem is beyond the scope of this paper.  It is important, however, to note that accurately reproducing the mass ratios of the terrestrial planets is a significant constraint for any successful numerical model of planetary formation.

Through dynamical modeling, we know chaos is prevalent in our solar system \citep{chaos4, chaos3, chaos2, laskar08}.  It is important to note the difference between ``stability" and ``chaos."  While a system without ``chaos" can generally be considered stable, a system with ``chaos" is not necessarily unstable \citep{milani92, chaos}.  As is convention in other works, in this paper ``chaos'' implies both a strong sensitivity of outcomes to specific initial conditions, and a high degree of mixing across all energetically accessible points in phase space \citep{chaos}  Conversely ``Instability'' is used to describe systems which experience specific dynamical effects such as ejections, collisions or excited eccentricities.  

The chaos in our solar system mostly affects the terrestrial planets, particularly Mercury, and can cause the system to destabilize over long periods of time.  \citet{laskar08} even shows a 1--2\% probability of Mercury's eccentricity being excited to a degree which would risk planetary collision in the next 5 Gyr.  What we still don't fully understand is whether these chaotic symptoms (highly excited eccentricities, close encounters and ejection) are an expected outcome of the planetary formation process as we presently understand it, or merely a quality of our particular solar system.  The work of \citet{laskar00} showed us that the outcomes of semi-analytic planetary formation models of our own solar system show symptoms of chaos, and are connected to the particular initial mass distribution which is chosen.  However these systems were formed without the presence of the gas giants, and planetesimal interactions were simplified to minimize computing time.  Perhaps our solar system is a rare outlier in the universe, with it's nearly stable, yet inherently chaotic system of orbits occurring by pure chance.  Of even greater interest, if it turns out that systems like our own are unlikely results of planetary formation, we may need to consider other mechanisms that can drive the terrestrial planets into their modern chaotic state.	

This work takes 145 systems of terrestrial planets formed in \citet{kaibcowan15} as a starting point.  The systems are broken into three ensembles.  The first set of 50 simulations, ``Circular Jupiter and Saturn" (cjs), are formed with Jupiter and Saturn on nearly circular (e\textless0.01) orbits, at their current semi-major axes.  The simulations use 100 self-interacting embryos on nearly circular and coplanar orbits between 0.5 and 4.0 AU, and 1000 smaller non-self-interacting planetesimals.  The smaller planetesimals interact with the larger bodies, but not with each other.  Additionally, the initial embryo spacing is uniform and embryo mass decreases with semi-major axis to yield an r\textsuperscript{-3/2} surface density profile.  The second ensemble (containing 46 integrations), ``Extra Eccentric Jupiter and Saturn" (eejs) evolve from the same initial embryo configuration as cjs, with Jupiter and Saturn initially on higher (e=0.1) eccentricity orbits.  The final batch of integrations (49 systems), ``Annulus" (ann), begin with Jupiter and Saturn in the same configuration as cjs, however no planetesimals are used.  400 Planetary embryos for ann are confined to a thin annulus between 0.7--1.0 AU, roughly representative of the conditions described following Jupiter's outward migration in the Grand Tack Model \citep{walsh11}.

After advancing each system to t=4.5 Gyr, we perform detailed 100 Myr simulations and probe multiple chaos indicators.  By careful analysis we aim to show whether chaotic systems naturally emerge from accretion models, and whether the source of the chaos is the same as has been shown for our own solar system.

\section{Methods}
\subsection{System Formation and Evolution}
We use the simulations modeling terrestrial planet formation in \citet{kaibcowan15} as a starting point for our current numerical work. In \citet{kaibcowan15}, all simulations are stopped after 200 Myrs of evolution, an integration time similar to previous studies of terrestrial planet formation \citep[e.g.][]{chambers01, raymond04, obrien06, walsh11}. Because we ultimately want to compare the dynamical state of our solar system (a 4.5 Gyr old planetary system) with the dynamical states of our simulated systems, we begin by integrating the systems from \citet{kaibcowan15} from $t=200$ Myr to $t=4.5$ Gyr. Since bodies can evolve onto crossing orbits and collide before $t=4.5$ Gyr, accurately handling close encounters between massive objects is essential. Thus, we use the MERCURY hybrid integrator \citep{chambers99} to integrate our systems up to $t=4.5$ Gyr. During these integrations, we use a 6-day timestep and remove bodies if their heliocentric distance exceeds 100 AU. Because we are unable to accurately integrate through very low pericenter passages, objects are also merged with the central star if their heliocentric distance falls below 0.1 AU. Though by no means ideal, the process of removing objects at 0.1 AU is commonplace in direct numerical models of planetary formation due to the limitations of the integrators used for such modeling.  \citet{chambers01} showed that this does not affect the ability to accurately form planets in the vicinity of the actual inner solar system, since objects crossing 0.1 AU must have very high eccentricities.  These excited objects interact weakly when encountering forming embryos due to their high relative velocity, and rarely contribute to embryo accretion.  It should be noted that many discovered exoplanetary systems have planets with semi-major axis interior to 0.1 AU.   However, we are not interested in studying such systems since we aim to draw parallels to our actual solar system.  The WHFAST integrator used in the second phase of this work (Section 2.2), however, can integrate the innermost planet to arbitrarily high eccentricities, so the 0.1 AU filter is no longer used.  Finally, to assess the dynamical chaos among planetary-mass bodies, any ``planetesimal'' particles (low-mass particles that do not gravitationally interact with each other) that still survive after 4.5 Gyrs are manually removed from the final system.

\subsection{Numerical Analysis}

\begin{figure}
\centering
\includegraphics[width=.45\textwidth]{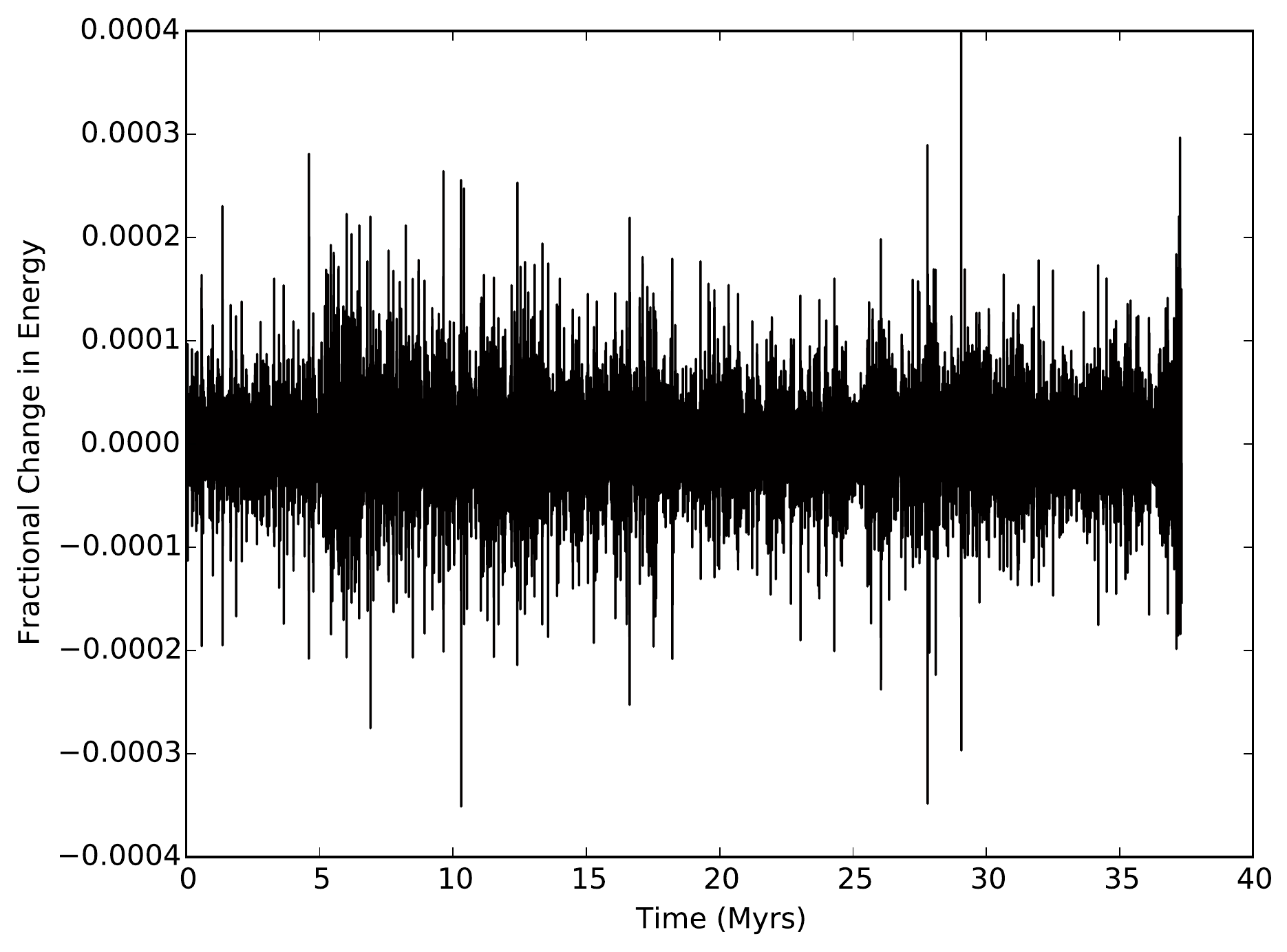}
\qquad
\includegraphics[width=.45\textwidth]{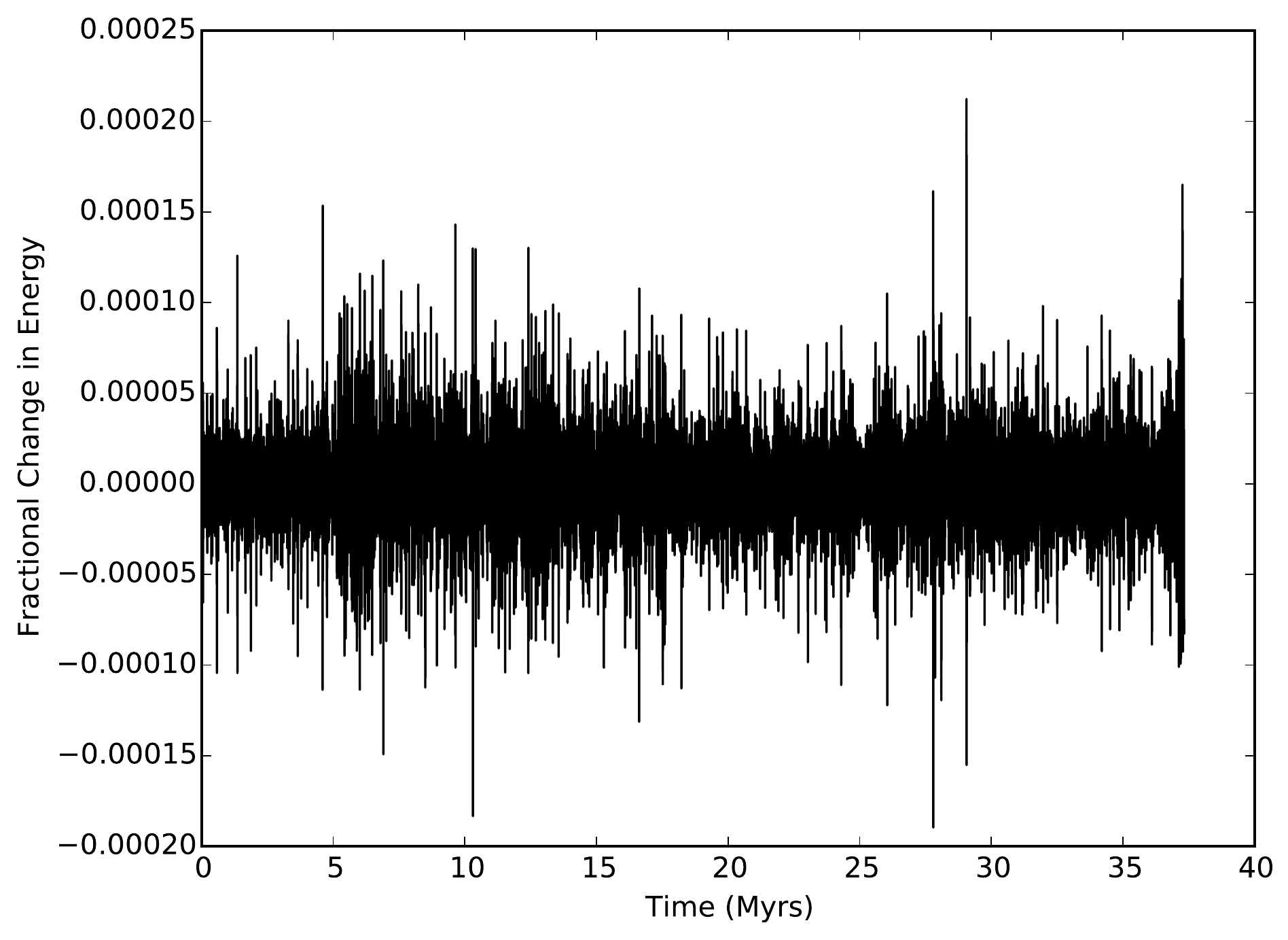}
\caption{$\bf{Upper}$: Fractional Energy ($\delta$E/E) change for the innermost planet in the system ann21, the body with the smallest pericenter ($q=0.136$ AU) in all of our systems.  $\bf{Lower}$: Fractional Energy change ($\delta$E/E) for all bodies in ann21.}
\label{fig:energy}
\end{figure}

\begin{figure}
\centering
\includegraphics[width=.5\textwidth]{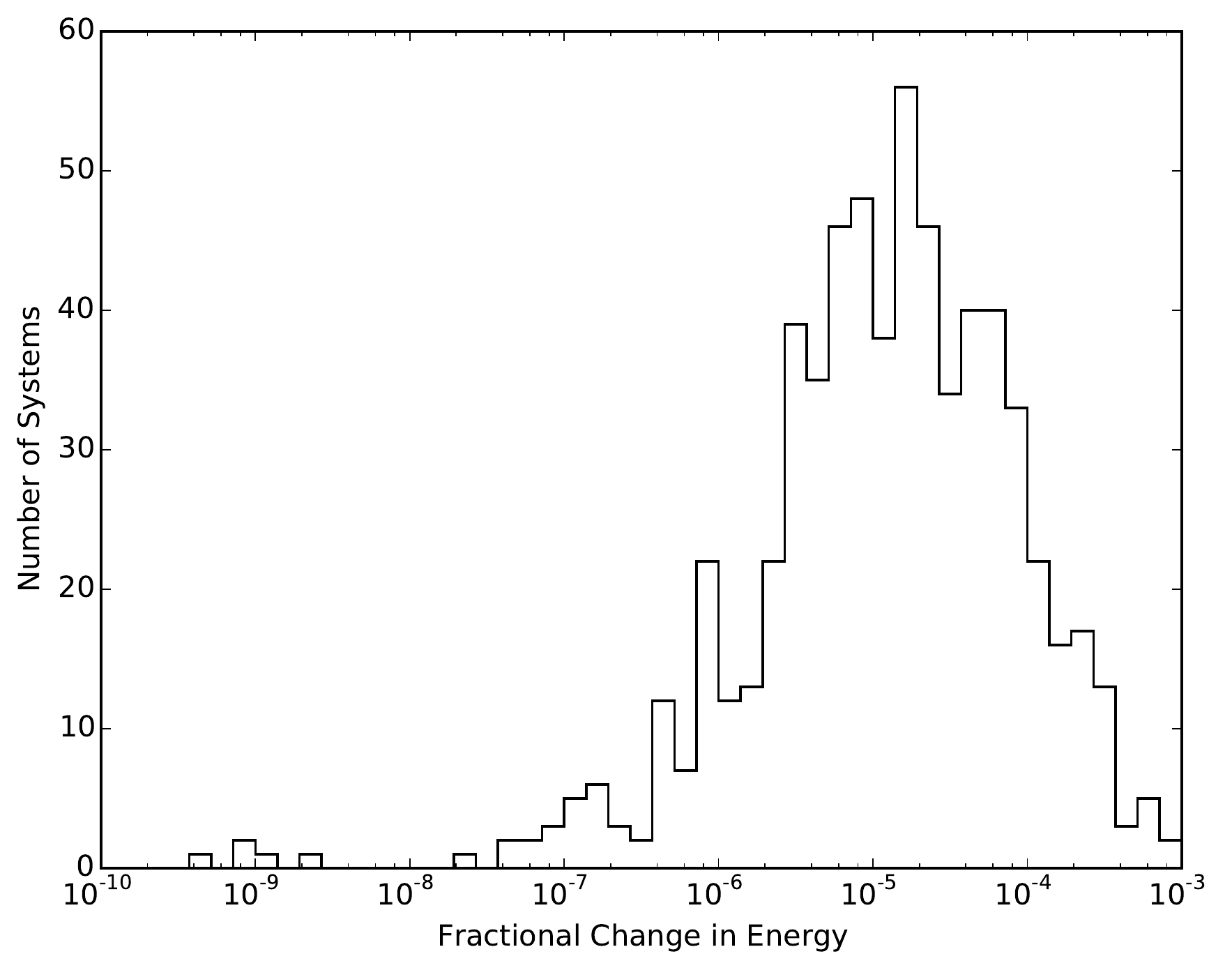}
\caption{Cumulative Distribution of the absolute value of Fractional Energy change ($\delta$E/E) over the duration of the simulation for all WHFAST integrations performed in this project.}
\label{fig:energy2}
\end{figure}

Numerical simulations for detailed analysis of the fully evolved systems are performed using the WHFAST integrator in the Python module Rebound \citep{rebound}.  WHFAST \citep{whfast} is a freely available, next generation Wisdom Holman symplectic integrator \citep{wh3,wh1,wh2} ideal for this project due to its reduction on the CPU hours required to accurately simulate systems of planets over long timescales.  WHFAST's reduction in error arising from Jacobi coordinate transformations, incorporation of the MEGNO (Mean Exponential Growth factor of Nearby Orbits) parameter, improved energy conservation error and tunable symplectic corrector up to order 11 motivate the integrator choice.  The accuracy of many mixed variable symplectic integration routines are degraded by integrating orbits through phases of high eccentricity and low pericenter.  For this reason, in Figure \ref{fig:energy} we plot the variation in energy from our simulation with the lowest pericenter ($q=0.136$ AU).  In the upper panel, we plot the energy of the innermost planet, since this should be fixed in the secular regime.  We see that energy variations stay well below one part in 10\textsuperscript{3}.  In the lower panel, we plot the fractional change in the total energy of this system.  Again, we find that energy variations rarely exceed one part in 10\textsuperscript{4}.  Finally, in Figure   \ref{fig:energy2} we show a histogram of the fractional energy change between the start and end of the integration for all of our simulations.  For the vast majority of systems, the fractional energy change is far less than one part in 10\textsuperscript{4}, and no simulations exceed 10\textsuperscript{3}.  Values in excess of 10\textsuperscript{4} are from simulations where the eccentricities of the giant planets were artificially inflated and the performance of the integrator is degraded by close encounters.  These systems with frequent close encounters are obviously chaotic, so our chaos determination is not affected.

MEGNO is the primary tool for identifying chaotic systems.  Introduced in \citet{megno2}, MEGNO represents the time averaged ratio of the derivative of the infinitesimal displacement of an arc of orbit in N-dimensional phase space to the infinitesimal displacement.  For quasi-periodic (stable) motion, MEGNO will converge to a value of 2 in the infinite limit.  For chaotic systems, however, MEGNO will diverge \citep{megno2}.  \citet{megno} showed that MEGNO is an extremely useful and accurate tool for detecting chaos.  Systems which are non-chaotic will maintain stable MEGNO values of $\sim$2 for the duration of the simulation, while chaotic systems diverge from 2.  For this project, systems which attained a maximum value of MEGNO $\geq$ 3.0 were classified as chaotic.

For use in certain analyses, the Lyapunov Timescale ($\tau_{L}$) is also output.  WHFAST calculates the inverse of $\tau_{L}$ by least squares fitting the time evolution of MEGNO \citep{whfast}.  Systems classified as chaotic tended to have a $\tau_{L}$ less than $\sim$10--100 Myr.  

Some simulations which quickly displayed chaos were terminated early to save computing time.  Terminating these simulations early did not affect the chaos determination since MEGNO had already clearly diverged, nor did shorter simulations affect follow on data analysis and reduction (such as the detection of resonances described in section 2.4).

Another tool we use to characterize the chaos in our systems is Angular Momentum Deficit (AMD) \citep{laskar97}.  AMD (equation \ref{eqn:amd}) measures the difference between the z-component of the angular momentum of a given system to that of a zero eccentricity, zero inclination system with the same masses and semi-major axes.  Evaluating the evolution of AMD over the duration of a simulation will probe whether angular momentum is being exchanged between giant planets and terrestrial planets as orbits excite and deexcite due to induced chaos \citep{amd}.  

\begin{equation}
	M_{z(def)} = \frac{\sum_{i}m_{i}\sqrt{a_{i}[1 - \sqrt{(1 - e_{i}^2)}\cos{i_{i}}]}} {\sum_{i}m_{i}\sqrt{a_{i}}} 
	\label{eqn:amd}
\end{equation} 

\subsection{Simulation Parameters}

Simulations are run for 100 Myr, with an integration timestep of 3.65 days.  Orbital data is output every 5000 years.  The order of the symplectic corrector is the order to which the symplectic correction term in the interaction Hamiltonian ($\epsilon$dt in \citet{whfast}) is expanded.  Here, we set this to order 1 (WHFAST allows for corrections up to order 11) \citep{whfast}.  10 sample systems were integrated at different corrector values in order to determine the lowest corrector order necessary to accurately detect chaos.  Additionally, a total of 16 1 Gyr simulations consisting of both chaotic and non-chaotic systems of 3, 4 and 5 terrestrial planets were performed to evaluate long-term behavior and verify the adequacy of 100 Myr runs.  Finally, 6 sets of 145 simulations are performed, results and findings for which are reported in section 3.  The 6 runs are summarized in Table \ref{table:sims}.

\begin{table}
\centering
\begin{tabular}{c c c c c}
\hline

Run & Run Time & Jupiter & Saturn & Outer Planet \\ Name & & & & eccentricity shift \\
\hline
1 & 100 Myr &  &  & unchanged\\
a & 100 Myr & removed & removed & unchanged\\
b & 100 Myr &  & removed & unchanged\\
c & 100 Myr &  &  & 150\%\\
d & 100 Myr &  &  & 200\%\\
e & 100 Myr/1 Gyr &  &  & + 0.05\\
\hline
\end{tabular}
\caption{Run 1 simulates each fully evolved 4.5 Gyr system discussed in section 2.1.  Runs a and b evaluate the effect of removing outer planets.  Runs c--e investigate the result of inflating the giant planet's eccentricities.  Run e is re-performed for cjs and ann systems using the MERCURY hybrid integrator for 1 Gyr in order to detect close encounters (see Section 3.2)}
\label{table:sims}
\end{table}

\subsection{Detecting Mean Motion Resonances}

A Mean Motion Resonance (MMR) occurs when the periods of orbital revolution of 2 bodies are in integer ratio to one another.  For a given MMR, the resonant angle will librate between 2 values \citep{resang}.  Many possible resonant angles exist.  For this paper, however, we only consider 4 of the more common planar resonant angles \citep{elliot05}.  Only planar resonances are considered because our systems typically have very low inclinations.

\begin{figure}
\centering
\includegraphics[width=0.39\textwidth]{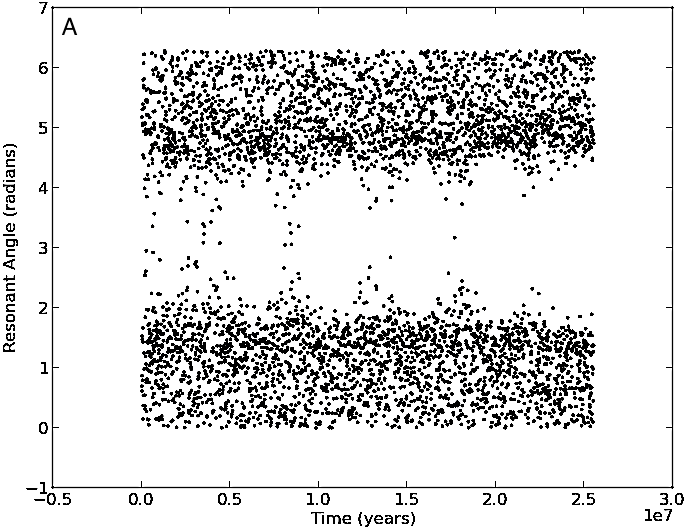}
\qquad
\includegraphics[width=0.39\textwidth]{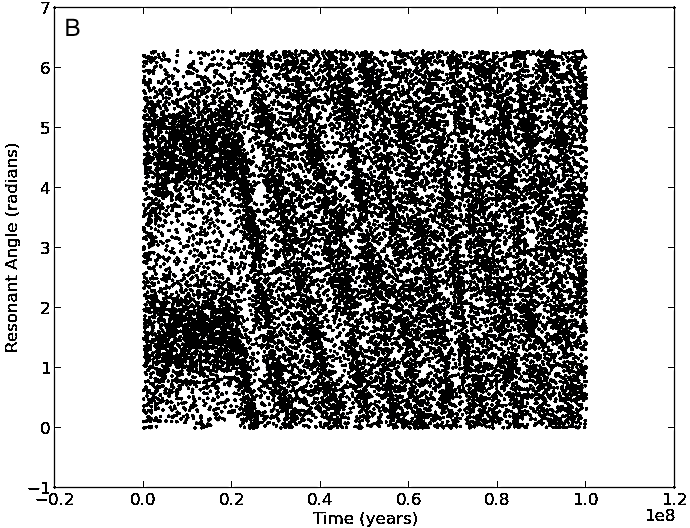}
\qquad
\includegraphics[width=0.39\textwidth]{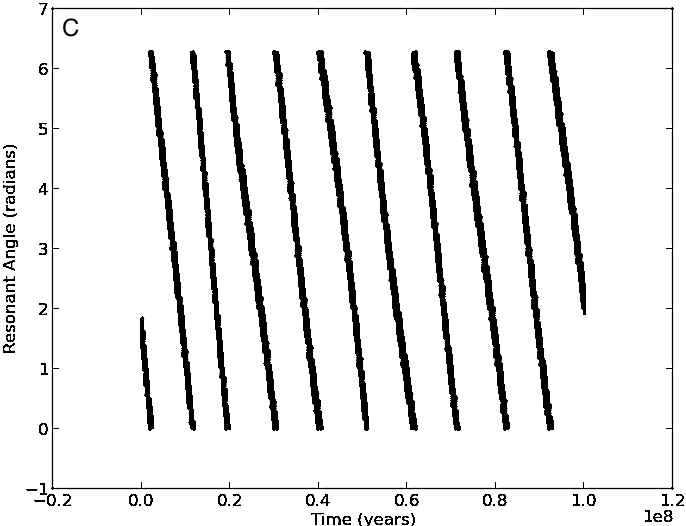}
\qquad
\includegraphics[width=0.39\textwidth]{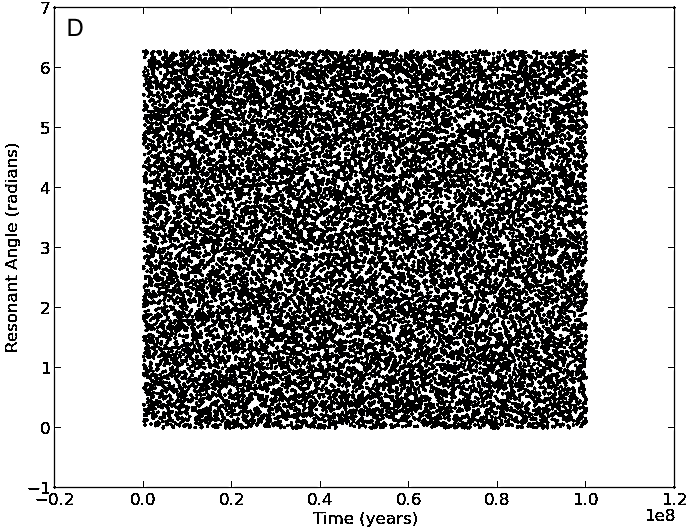}
\caption{$\bf{A}$ shows clear libration between 2 resonant angles.  These 2 terrestrial planets are locked in a 2:1 MMR for the duration of the simulation.  $\bf{B}$: depicts occasional libration where 2 rocky planets are going in and out of a 5:3 MMR.  $\bf{C}$ shows slow circulation of a resonant angle for 2 inner planets.  These objects are close to a 2:1 MMR. Finally, $\bf{D}$ is an example of a uniform distribution of resonant angles where the objects are not in a MMR.}
\label{fig:mmrs}
\end{figure}

To detect MMRs, the average keplerian period is calculated for all bodies in the simulation.  4 resonant angles are calculated for all sets of bodies with period ratios within 5\% of a given integer ratio.  21 different MMRs (all possible permutations of integer ratios between 2:1 and 8:7) are checked for.  Using a Komolgorov-Smirnov test, each resulting time-resonant angle distribution (e.g. figure \ref{fig:mmrs}) is compared to a uniform distribution (e.g. figure \ref{fig:mmrs}d), yielding a p-value.  All distributions with p-values less than 0.01 are evaluated by eye for libration.  Figure \ref{fig:mmrs} shows 4 different example distributions and their classification.

\section{Results}

\subsection{System Evolution Beyond 200 Myrs}

In \citet{kaibcowan15}, systems of terrestrial planets were generated via simulations of terrestrial planet accretion. These simulations were terminated after 200 Myrs of evolution, as each simulation had evolved into a system dominated by 1--6 terrestrial planet-mass bodies. Terminating accretion simulations after 200--400 Myrs of system evolution is common practice since the great majority of accretion events occur well before these final times are reached. However, it remains unknown how these newly formed systems evolve over the next several Gyrs. Do planetesimals and embryos naturally accrete into indefinitely stable configurations of terrestrial planets? Or are the systems that arise from terrestrial planet accretion often only marginally stable, with major instabilities occurring hundreds of Myrs or Gyrs after formation?

\begin{figure}
\centering
\includegraphics[width=.5\textwidth]{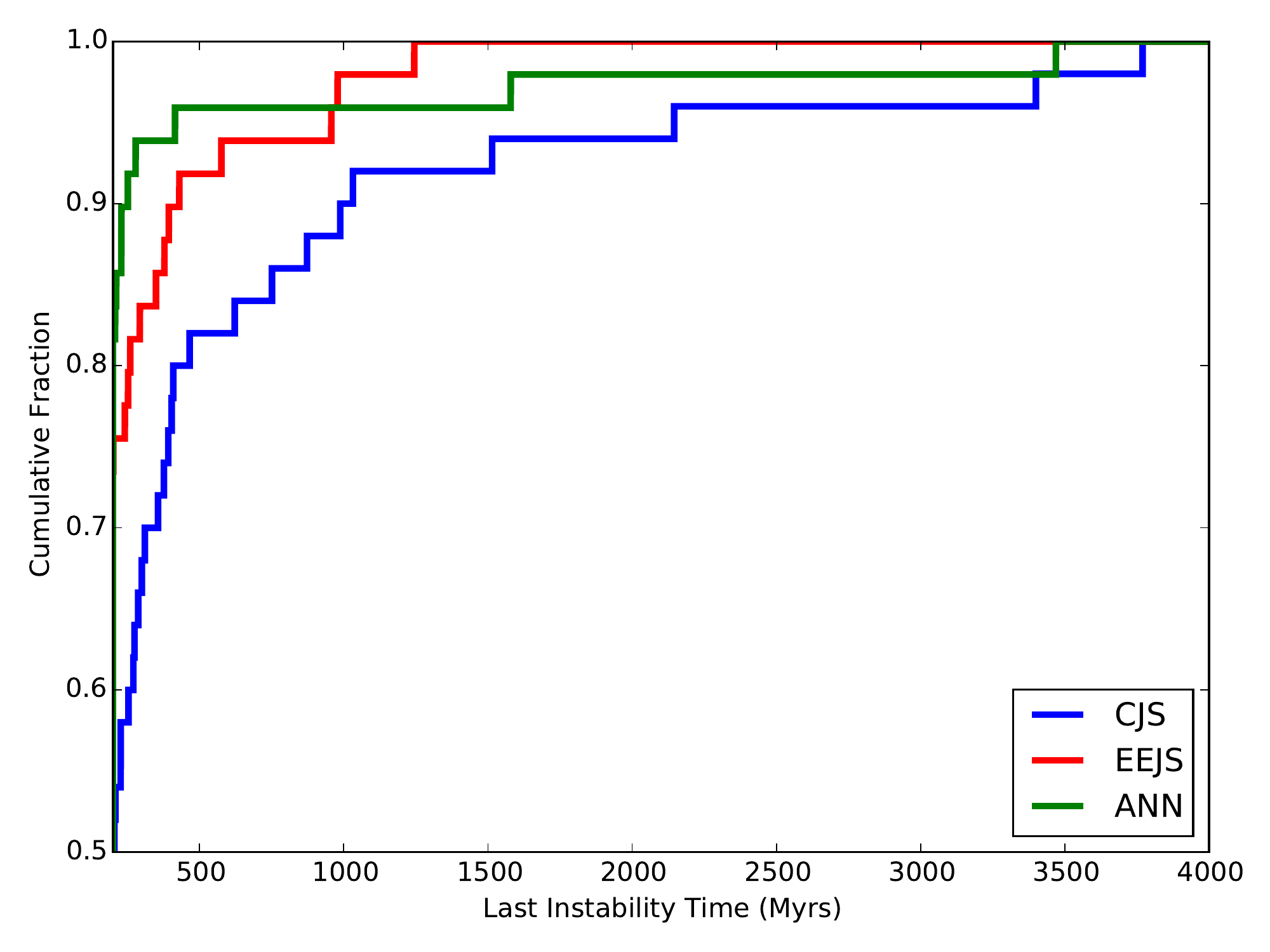}
\caption{Cumulative distribution of the last times at which systems lose a planet more massive than 0.055 M$_{\oplus}$ via collision or ejection. The distributions for cjs, eejs, and ann are shown with the blue, red, and green lines, respectively.}
\label{fig:lasttime}
\end{figure}

To begin answering this question, we take the systems from \citet{kaibcowan15} and integrate them for another 4.3 Gyrs with MERCURY. In Figure~\ref{fig:lasttime}, we show the cumulative distributions of times at which these systems lose their last terrestrial planet mass body ($m>$ 0.055 M$_{\oplus}$). These planets can be lost via collision with a larger planet, collision with the Sun, or ejection from the system ($r>100$ AU). We find that there are many systems that undergo substantial dynamical evolution after their first 200 Myrs. As Figure~\ref{fig:lasttime} shows, between 20 and 50\% of systems lose at least 1 planet after $t=200$ Myrs. This fraction varies with the simulation batch.  Systems in the cjs set are the most likely to lose planets at late times. This is likely due to the fact that these systems often form planets well beyond 2 AU \citep{ray09a}, where dynamical timescales are longer and planets require a longer time period to undergo ejections or final collisions compared to those at $\sim$1 AU. In eejs simulations, a smaller fraction of systems ($\sim$25\%) lose a planet after their first 200 Myrs of evolution. In these systems, the effects of an eccentric Jupiter and Saturn greatly deplete the mass orbiting beyond 1.5--2 AU  \citep{ray09a}, and this absence of more distant material may explain the decrease in late instabilities. Finally, the conditions are even more extreme in the ann simulations, where the initial planetesimal region is truncated at 1 AU. These simulations have the lowest rate of late ($t>200$ Myrs) instabilities at 18\%. 

It should also be noted that some systems lose planets at extremely late times. 8 out of 150 systems ($\sim$5\%) lose planets after $t=1$ Gyr. 5 of these systems are from cjs, while eejs and ann yield 1 and 2 systems, respectively. This small, yet non-negligible fraction of systems undergoing late instabilities may help explain the existence of transient hot dust around older main sequence stars \citep{wyatt07}. These very late instabilities in our systems occur even though the orbits of Jupiter and Saturn are effectively fixed for the entire integration. The rate of instabilities would likely be significantly higher if the orbits of the gas giants evolved substantially over time \citep{bras09, agnorlin12, bras13, kaibcham16}. 

\begin{figure}
\centering
\includegraphics[width=.5\textwidth]{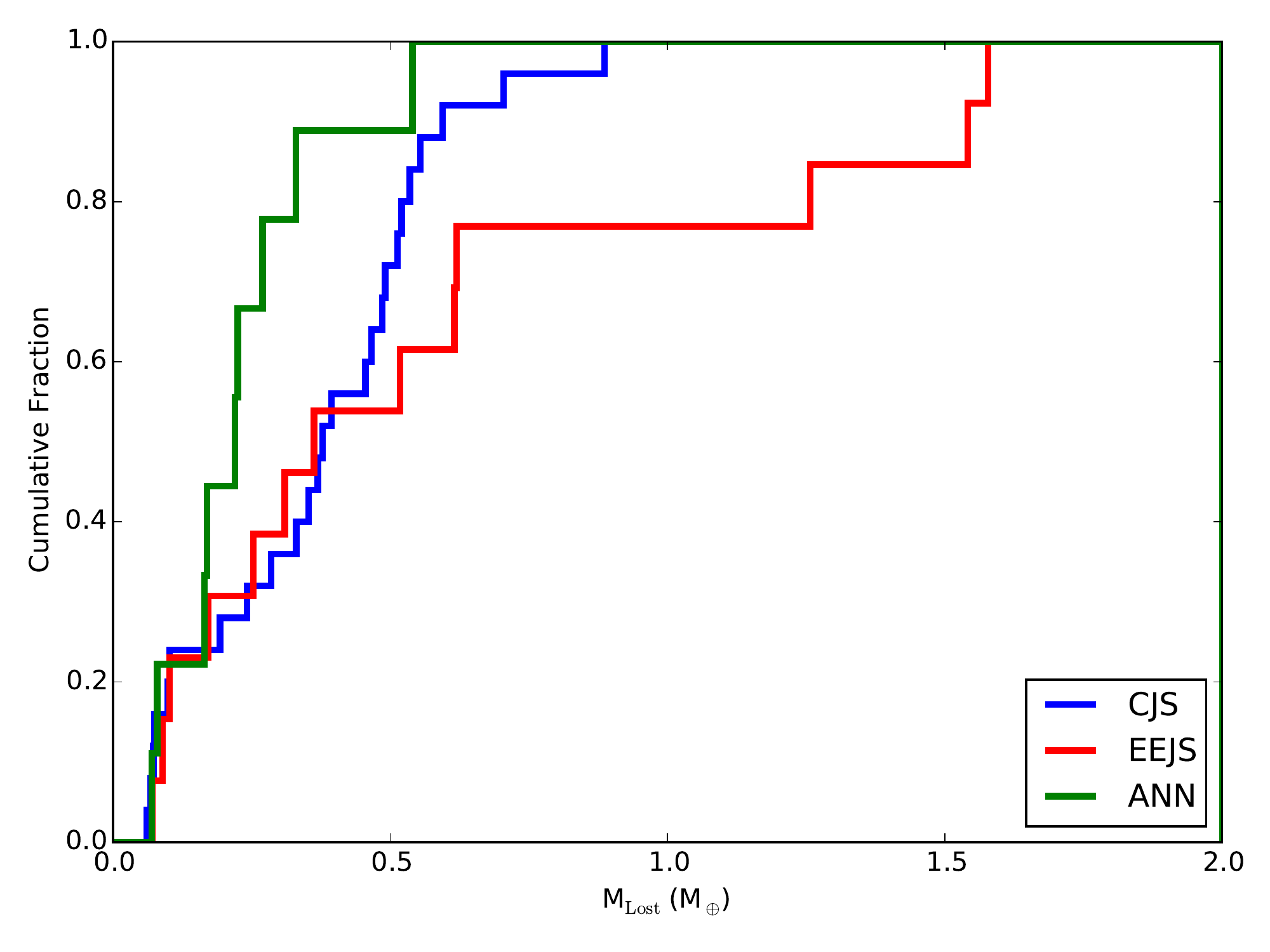}
\caption{For systems that lose a planet after $t=200$ Myrs (more massive than 0.055 M$_{\oplus}$), the distribution of masses for the last lost planet is shown. The distributions for cjs, eejs, and ann are shown with the blue, red, and green lines, respectively.}
\label{fig:lastmass}
\end{figure}

In Figure~\ref{fig:lastmass}, we look at the mass distributions for the last planet lost from each system with an instability after $t=200$ Myrs. In general, we see that the last planets lost from systems with late instabilities have masses below $\sim$0.5 M$_{\oplus}$. This is not surprising, since during an instability event it is typical for the smallest planets to be driven to the highest eccentricities, resulting in their collision or ejection \citep{rasford96, chatterjee08, jurtre08, ray09b}. However, not all systems abide by this. In particular, 3 of the 13 eejs systems that undergo late instabilities lose planets with masses well over 1 M$_{\oplus}$. 

This suggests there may be a different instability mechanism in eejs systems. Indeed, when we look at {\it how} the last planets are lost from eejs systems, we find that 10 of the 13 systems with late instabilities lose their planets via collision with the Sun. This contrasts strongly with the cjs and ann systems, where there is only one instance of a planet-Sun collision among the 34 systems that have late instabilities. Moreover, there are 4 eejs systems with only 1 planet at $t=200$ Myrs, which go on to have a planet-Sun collision before $t=4.5$ Gyrs. In these cases, the gas giants are clearly driving instabilities. This is not surprising, since the heightened eccentricities of Jupiter and Saturn will enhance the secular and resonant perturbations they impart on the terrestrial planets. When interactions between a gas giant and the terrestrial planets are the main driver of an instability, the relative masses of the terrestrial planets lose their significance because they are all so small relative to the gas giants. This allows for more massive planets to be lost from these systems.

\begin{figure}
\centering
\includegraphics[width=.5\textwidth]{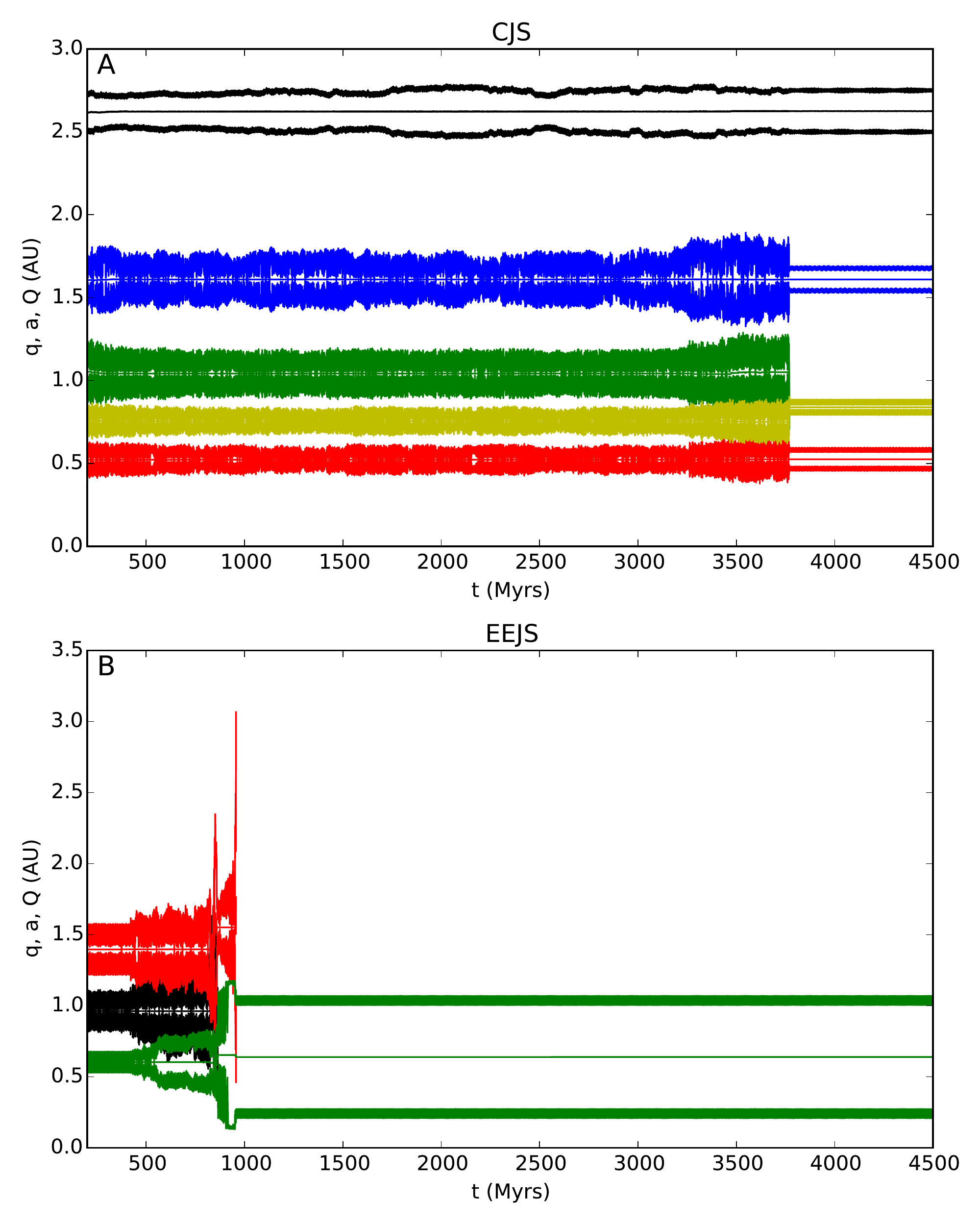}
\caption{{\bf A:} Time evolution of a 5-planet system from the cjs simulation batch. The pericenter, apocenter, and semi-major axis of each terrestrial planet is shown vs time. {\bf B:} Time evolution of a 3-planet system from the eejs simulation batch. The pericenter, apocenter, and semi-major axis of each terrestrial planet is shown vs time.}
\label{fig:examp}
\end{figure}

Figure~\ref{fig:examp}A shows an example of an instability within a cjs system. In this case, a system of 5 terrestrial planets are orbiting at virtually fixed semi-major axes for 3.8 Gyrs when an instability develops between the inner 3 planets. The second and third planets collide and the resulting 4-planet system finishes the simulation with smaller orbital eccentricities than it began with. On the other hand, the evolution of an eejs system is shown in Figure~\ref{fig:examp}B. Here we see the eccentricities of 3 relatively well separated planets driven up around 400--500 Myrs, leading to a collision between the second and third planets. After the collision, the outermost planet's eccentricity is again quickly excited and eventually approaches 0.8. Shortly after this point, the planet collides with the Sun (after a scattering event with the inner planet). While the detailed dynamics of this system are undoubtedly complex, the behavior is clearly different from that of the cjs system, and is almost certainly a consequence of the enhanced gas giant perturbations produced from their increased eccentricities.

\begin{figure}
\centering
\includegraphics[width=.5\textwidth]{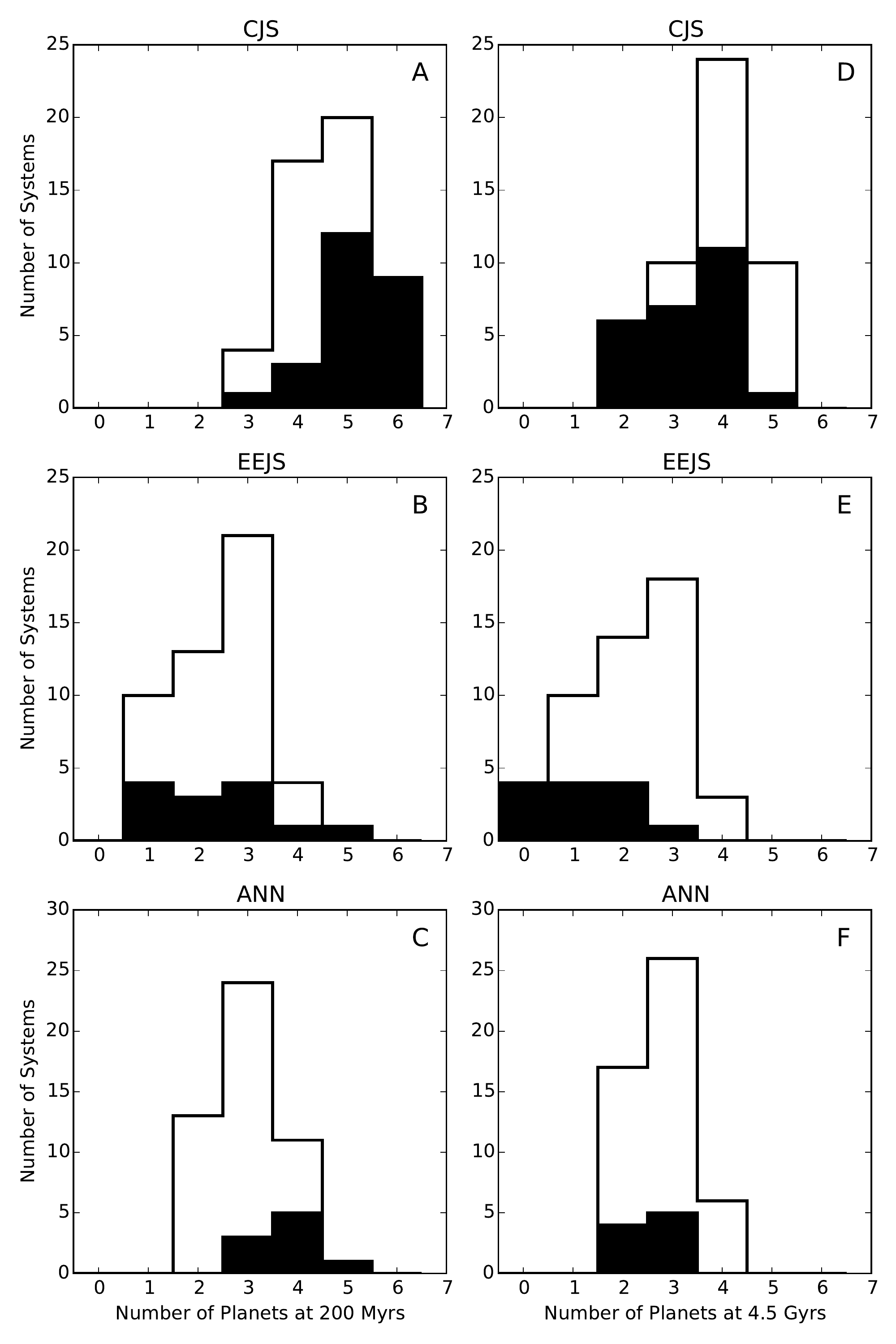}
\caption{{\bf A--C:} The distribution of the number of planets in each system at $t=200$ Myrs are shown for our cjs, eejs, and ann simulation batches in panels A, B, and C, respectively. The unfilled histograms show the distribution for all systems, and the filled histograms shown the distribution for systems that lose at least 1 planet more massive than 0.055 M$_{\oplus}$ after $t=200$ Myrs. {\bf D--F:} The distribution of the number of planets in each system at $t=4.5$ Gyrs are shown for our cjs, eejs, and ann simulation batches in panels D, E, and F, respectively. The unfilled histograms show the distribution for all systems, and the filled histograms shown the distribution for systems that lose at least 1 planet more massive than 0.055 M$_{\oplus}$ after $t=200$ Myrs.}
\label{fig:npl}
\end{figure}

We also study how the properties of systems with late instabilities differ from systems that do not lose planet-mass bodies after $t=200$ Myrs. In Figure~\ref{fig:npl} we look at the number of planets that each system has. In panels A--C, we see that after 200 Myrs of evolution, cjs systems typically have 4--6 planet-mass bodies (an average of 4.68 planets per system). This is significantly higher than eejs and ann systems, which have an average of 2.45 and 3.00 planets per system, respectively. The differences can largely be attributed to the lack of distant planets in these systems, owing to their initial conditions. Panels A--C also show which systems go on to lose planets at later times. For cjs and ann simulations, these systems tend to have more planets than the overall distribution. In contrast, no such trend is seen among eejs systems. Regardless of planet number, the eejs systems all seem to have roughly the same probability of losing a planet at late times. This is again a symptom of the gas giants driving instabilities within these systems, unlike the cjs and ann systems, where interactions between terrestrial planets play a larger role in late instabilities. Finally, panels D--F show the distributions of planets per system after 4.5 Gyrs of evolution. At the end of our integrations, the cjs, eejs, and ann systems have an average of 3.76, 2.12, and 2.78 planets per system respectively. For all of our simulation batches, we see that systems with late instabilities tend to have lower numbers of planets than the overall distribution of systems. Thus, in the case of cjs and ann systems, late instabilities tend to transform systems with relatively high numbers of planets into systems with relatively few planets. We also note that 4 eejs systems finish with no terrestrial planets whatsoever. 

\begin{figure}
\centering
\includegraphics[width=.5\textwidth]{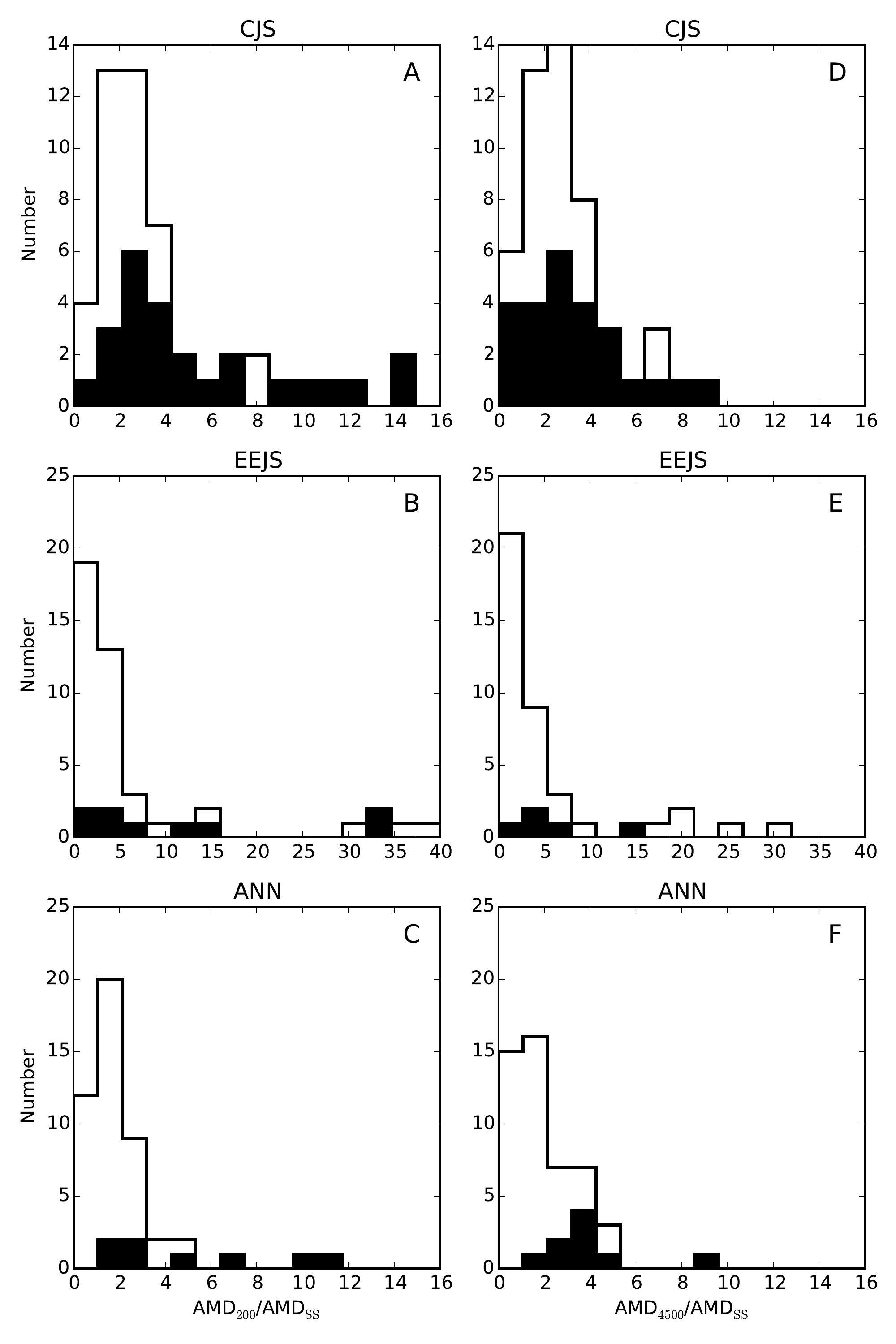}
\caption{{\bf A--C:} The distribution of terrestrial AMD values (normalized to the solar system's modern terrestrial AMD) at $t=200$ Myrs are shown for our cjs, eejs, and ann simulation batches in panels A, B, and C, respectively. The unfilled histograms show the AMD values for all systems, and the filled histograms shown the AMD values for systems that lose at least 1 planet more massive than 0.055 M$_{\oplus}$ after $t=200$ Myrs. {\bf D--F:} The distribution of terrestrial AMD values (normalized to the solar system's modern terrestrial AMD) at $t=4.5$ Gyrs are shown for our cjs, eejs, and ann simulation batches in panels A, B, and C, respectively. The unfilled histograms show the AMD values for all systems, and the filled histograms shown the AMD values for systems that lose at least 1 planet more massive than 0.055 M$_{\oplus}$ after $t=200$ Myrs.}
\label{fig:amd}
\end{figure}

Finally, we show the AMD of each of our terrestrial planet systems at $t=200$ Myrs and $t=4.5$ Gyrs in Figure~\ref{fig:amd}. Panels A--C show our systems' AMD distribution at $t=200$ Myrs. For each of our simulation batches, the median AMD is greater than the solar system's value. Our cjs, eejs and ann simulations have median AMD values of 2.9, 4.0, and 1.7 times the value of the modern inner solar system. Again, we also show the AMD distributions for systems that go on to have late instabilities. These systems tend to have larger AMD values. For the cjs, eejs and ann systems that undergo late instabilities the median AMD values at $t=200$ Myrs are 4.1, 14, and 4.3 times the solar system's AMD, respectively. Interestingly, though, panels D--F demonstrate that these systems are not always destined to maintain a relatively large AMD. Systems in the cjs and ann batches that undergo late instabilities have median AMD values of 2.7 and 3.7 times the value of the solar system after 4.5 Gyrs of evolution, respectively. Thus, a late instability does not necessarily increase the AMD of the system, and in some situations can result in moderate decreases. On the other hand, in eejs systems, the excited orbits of Jupiter and Saturn continue to wreak havoc on the terrestrial planets. Systems that experience late instabilities have a median AMD of 132 times that of the solar system!

\subsection{Prevalence of Chaos}

\begin{figure}
\centering
\includegraphics[width=.5\textwidth]{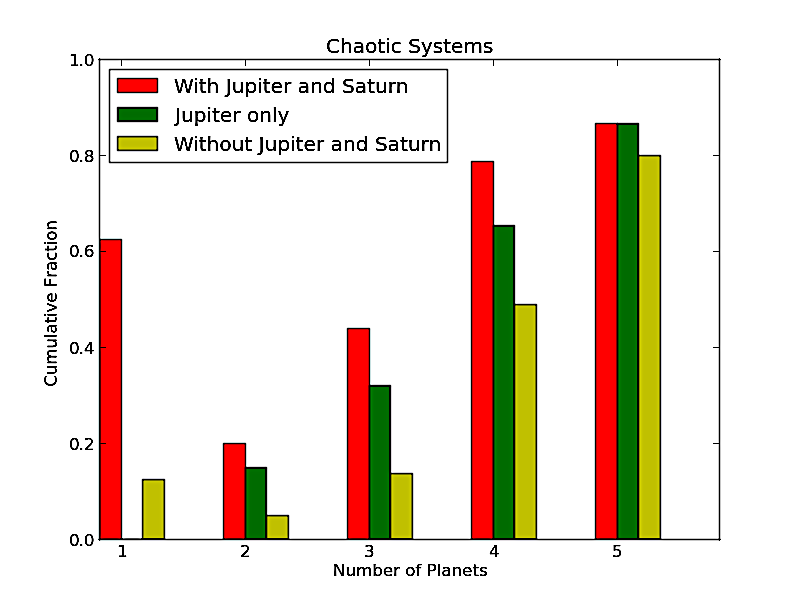}
\caption{Comparison of the prevalence of chaos for runs 1, a and b.}
\label{fig:chaos}
\end{figure}

A selection of results from our simulations are provided in Appendix A (Table \ref{table:data}).  $\tau_{L}$ and MEGNO are listed for run 1 for all systems.  Additionally, we provide our chaos determination (yes or no) for runs 1, a and b, as well as MMRs detected for run 1.  Figure \ref{fig:chaos} compares the fraction of all systems which are chaotic between runs 1, a and b.  We find that removing Jupiter and Saturn has the greatest effect on reducing chaos in our systems.  In general, $\sim50$\% of systems exhibit some form of chaos, when Saturn is removed only $\sim40$\% of systems are chaotic and when Jupiter is removed that number is only $\sim20$\%.  This indicates that the chaos in most of our systems is likely driven by perturbations from Jupiter.  In fact, when Jupiter was removed, all systems but 1 had $\tau_{L}$'s which either increase, or are within 1.5 orders of magnitude of the original value. 

\begin{figure}[h]
\centering
\includegraphics[width=.5\textwidth]{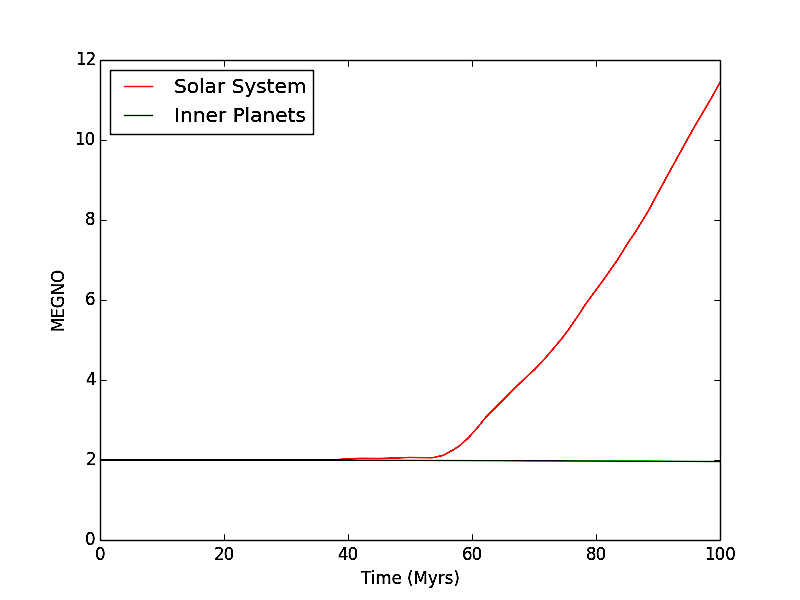}
\caption{This figure compares the behavior of MEGNO in two separate simulations of our own solar system, run using the same settings described in section 2.3.  With the giant planets in place (red line), the system is chaotic, and when they are removed (green line) the chaos disappears. Though Jupiter is the source of the chaos in our own solar system, when only Saturn, Uranus and Neptune are removed the system is non-chaotic because the precession of Jupiter's orbit due to the outer planets stops.  This result is different than the vast majority of the systems in this paper, which remain chaotic when only Saturn is removed.}
\label{fig:sol}
\end{figure}

Figure \ref{fig:chaos} also clearly shows the disparity of chaos between systems with different numbers of terrestrial planets.  This is most pronounced in 5-planet configurations, where only 2 such systems are free of chaos with the outer planets in place, and only 3 when they are removed.  Having 4 terrestrial planets may not to be a significant source of chaos in our own solar system.  This can be seen in figure \ref{fig:sol}, which provides MEGNO plots for the solar system with and without the 4 outer planets.  In fact, the solar system may better be described as a 3-planet configuration when compared to our results. Most 4 and 5-planet systems in our study differ greatly from our own since they typically contain only planets with masses comparable to Earth and Venus (see Table \ref{table:cjs1013} for 2 such 5-planet examples).  Mars analogues are rare in our systems, and Mercury sized planets are almost non-existent.  If we consider our solar system a 3 terrestrial planet arrangement, it's inherent chaos fits in well with our results; where about half of systems show chaos with the giant planets in place, and only around 1 in 6 when they are removed.  A shortcoming of this comparison is that many studies have shown that Mercury is a very important source of the chaos in our own solar system \citep{laskar08, laskar09}.  However, when we integrate the solar system without Mercury, the system is still chaotic.  Therefore, though the actual solar system does match our results, this comparison is limited by the fact that present models of the terrestrial planet formation systematically fail to produce Mercury analogs.  

MMRs between planets are common features in many of our chaotic systems, implying that they are often important sources of the dynamical chaos.  We detect 365 MMRs among all simulations in this phase of the project, 82\% of which occur in chaotic systems.  Further analysis shows that the MMRs which do occur in non-chaotic systems tend to be of higher order between smaller terrestrial planets.  It should be noted that the vast majority of these MMRs are intermittent, and last only a fraction of the entire simulation duration.

\begin{figure}
\centering
\includegraphics[width=.5\textwidth]{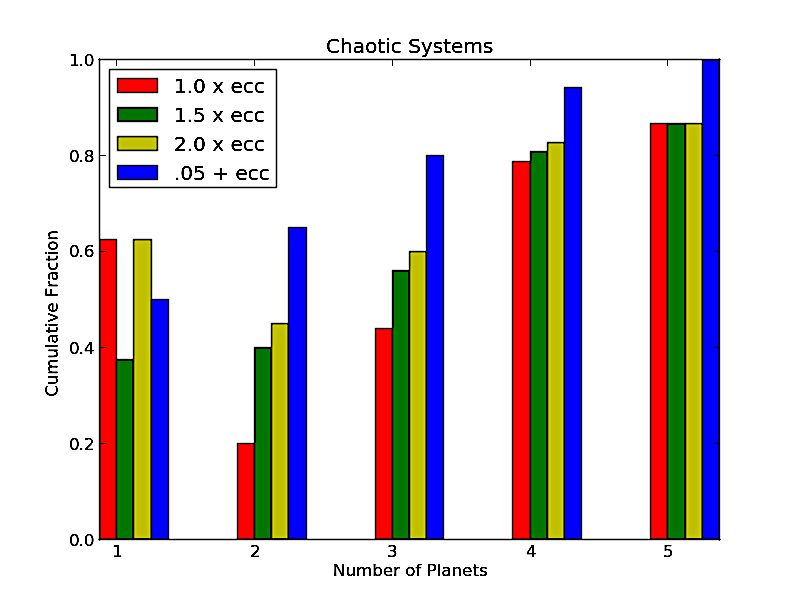}
\caption{Comparison of the prevalence of chaos for runs 1, c, d and e.}
\label{fig:ecc}
\end{figure}

Figure \ref{fig:ecc} shows the fraction of systems which are chaotic in runs 1, c, d and e.  It is clear that an eccentric Jupiter and Saturn can quickly introduce chaos to an otherwise non-chaotic system.   One interesting result from this batch of simulations is that when the eccentricities of the outer planets are inflated, the likelihood of a 5:2 MMR between Jupiter and Saturn developing increases.  In almost all systems, this resonant perturbation introduces chaos, and can possibly destabilize the system.  Since systems labeled cjs were formed with the giant planets on near circular orbits, simply multiplying the already low eccentricity by 1.5 or 2 was not enough to produce a noticeable effect.  For this reason, run e was performed using a step increase of 0.05.  There is a clear parallel between the results of this scenario and a Nice Model instability \citep{gomes05, mor05, tsi05}, where the outer planets rapidly transition from nearly circular to relatively eccentric orbits.  

\begin{table}
\centering
\begin{tabular}{c c c c c c c}
\hline
     system  & 0 & 1 & 2 & 3 & 4 & 5\\
\hline
                ann   &   71\% &    12\% & 2\% & 12\% &  0\% & 4\%\\
                cjs   &   48\% &   18\%  & 18\% & 14\% &  2\% & 0\%\\
\hline
\end{tabular}
\caption{The percentage of ann and cjs systems which lost a given number of planets when integrated with inflated giant planet eccentricities in run e for 1 Gyr.}
\label{table:ecc3}
\end{table}

\begin{figure}
\centering
\includegraphics[width=.5\textwidth]{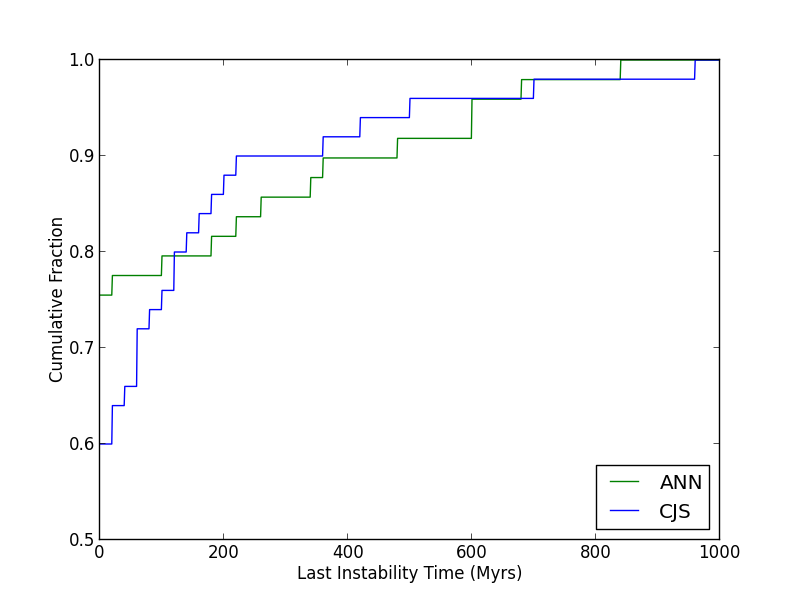}
\caption{Cumulative distribution of the last times at which systems' with inflated eccentricities lose a planet more massive than 0.055 M$_{\oplus}$ via collision or ejection. The distributions for cjs and ann are shown with the blue and green lines, respectively.}
\label{fig:ecc2}
\end{figure}

To further probe this effect, we repeat run e for cjs and ann systems using the MERCURY hybrid integrator in order to accurately detect collisions and ejections.  Systems are integrated for 1 Gyr using simulation parameters similar to those discussed in section 2.1.  We find instabilities are relatively common in these systems.  29\% of ann systems and 52\% of cjs systems lose one or more planets over the 1 Gyr integration.  Table \ref{table:ecc3} shows the percentage of systems which lose a given number of terrestrial planets.  In fact, the resulting systems are quite similar to those produced after integrating the eejs batch to the current epoch.  A small fraction of systems lose all inner planets, and some can have instabilities occur very late in the simulations (Figure \ref{fig:ecc2}).  Overall we show that an event similar to the Nice Model scenario, where the Giant planets eccentricities quickly inflate, can result in a non-negligible probability of inner planet loss.

If we again classify our solar system as a 3 terrestrial planet system, we can draw further parallels between the results of such cjs run e configurations and the Nice Model instability, since Jupiter and Saturn begin on near circular orbits.  In run e, 6 out of 8 such systems were chaotic, as compared to only 2 out of 8 in run 1.  Indeed, we see that even a perturbation in the giant planet's eccentricities of 0.05 is successful in rapidly making a system chaotic.  In fact, when our own solar system is integrated with the outer planets on circular (e\textless0.001, i$\sim$0) orbits, the chaos disappears.

\subsection{Angular Momentum Deficit and Last Loss Analysis}
For our systems, we evaluate the difference between the average AMD of the inner planets over the first and last 3 Myr of our 100 Myr simulations.  Taking the average removes the contributions from periodic forcing in the AMD from Jupiter and Saturn.  We find a weak trend for chaotic 4 and 5 planet systems to have larger changes in AMD over the duration of the simulation than their non-chaotic counterparts.  Of the 13 systems which had total changes in AMD greater than the actual solar system's value, 9 were classified as chaotic.  The largest outlier, a non-chaotic eejs 2 planet system (eejs25), is discussed further in section 3.5.1.  We also search for any correlation between the time and mass of the last object ($m>$ 0.055 M$_{\oplus}$) lost, and the chaos of a system.  Though we show that some unstable, chaotic systems can stabilize after losing a planetary mass body, we are unable to identify any conclusive trends as to whether a late instability will shape the ultimate chaotic state of a system.

\subsection{Mass Concentration Statistic and Center of Mass Analysis}

To evaluate the degree to which mass is concentrated at a given distance away from the central star, we utilize a mass concentration statistic ($S_{c}$) \citep{amd}:

\begin{equation}
	S_{c} = MAX\bigg(\frac{\sum_{i}m_{i}} {\sum_{i}m_{i}[\log_{10}(\frac{a}{a_{i}})]^2}\bigg) 
	\label{eqn:sc}
\end{equation}

\begin{figure}
\centering
\includegraphics[width=.5\textwidth]{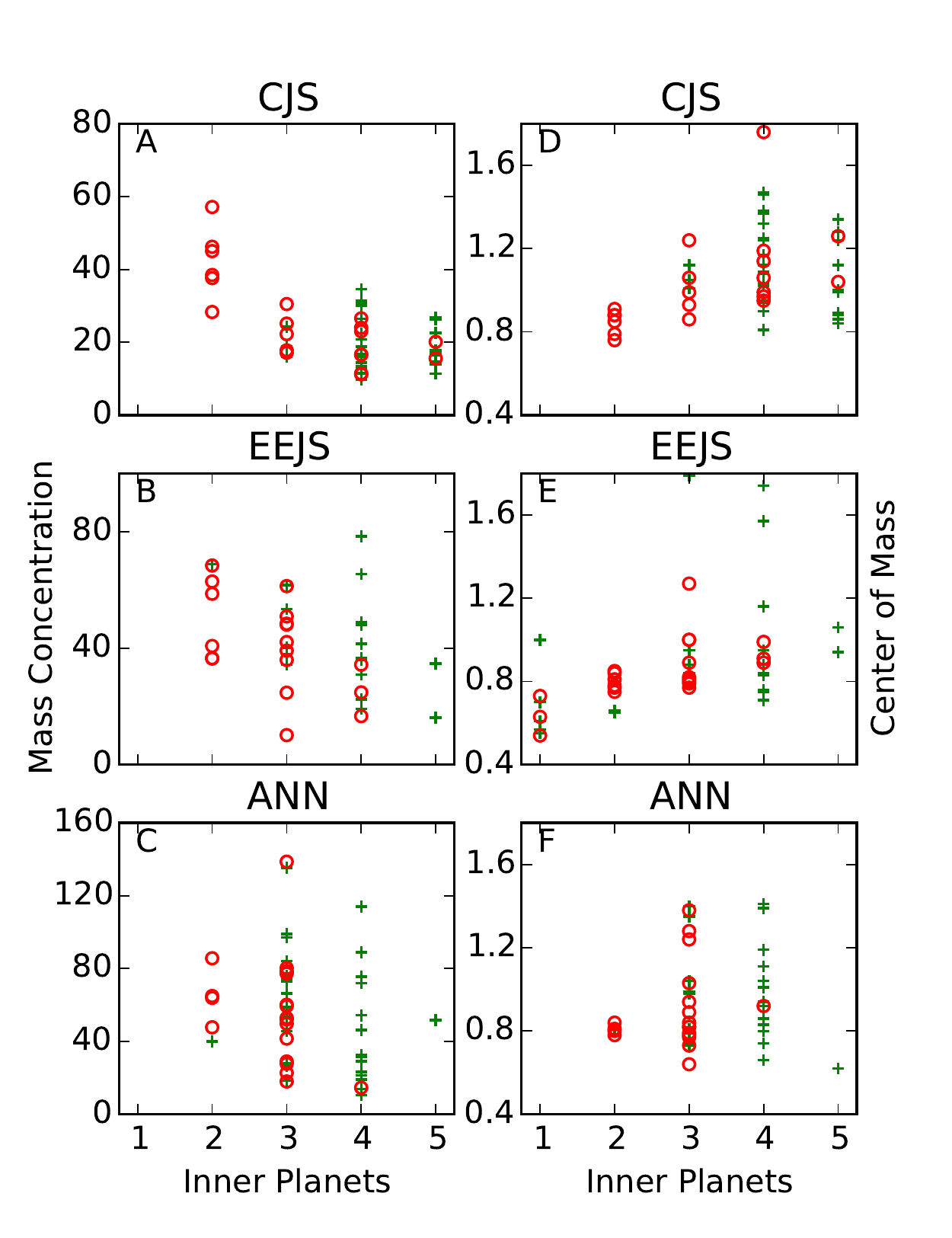}
\caption{{\bf A--C:} $S_{c}$'s affect on chaos for simulation batches cjs, eejs and ann (A, B and C respectively).  $S_{c}$'s for 1-planet systems are infinite, and therefore omitted in these graphs.  {\bf D--F:} Relationship between center of mass and chaos in simulation batches cjs, eejs and ann (D, E and F respectively).  Chaotic systems are designated by green plus signs and non-chaotic systems are represented by red circles.}
\label{fig:sc}
\end{figure}

The expression in parenthesis in \eqref{eqn:sc} is essentially the level of mass concentration at any point as a function of semi-major axis.  \citet{amd} utilizes a logarithm in the equation since, in our own solar system, the semi-major axes of the planets lie spaced in rough geometric series (the famous Titus-Bode law).  $S_{c}$ is the maximum value of the mass concentration function.  A system where most of the mass is concentrated in a single, massive planet would have a very steep mass concentration curve, and a high value of $S_{c}$.  A system of multiple planets with the same mass would have a smoother curve and yield a lower $S_{c}$.  As a point of reference, the $S_{c}$ value of the solar system's 4 inner planets, where most of the mass is concentrated in Venus and Earth, is 90.  $S_{c}$ values are provided in the same format as AMD values in figure \ref{fig:sc}.  A general, weak correlation can be seen between chaotic systems and slightly higher values of $S_{c}$, however this trend is not very conclusive.
We also provide the center of mass for each system of terrestrial planets in figure \ref{fig:sc}.  A clear trend is visible where non-chaotic systems tend to have a center of mass between $\sim$0.8--1.2 AU (the value being slightly greater as number of planets increases).  In general, the more mass is concentrated closer to the central star, or closer to Jupiter, the greater the likelihood of chaos developing. This is likely related to Jupiter's role in introducing chaos to systems.  This trend is true for all three simulation subsets, but particularly strong in the cjs and eejs batches.

\subsection{Systems of Particular Interest}

\subsubsection{eejs25}

\begin{figure}
\centering
\includegraphics[width=0.5\textwidth]{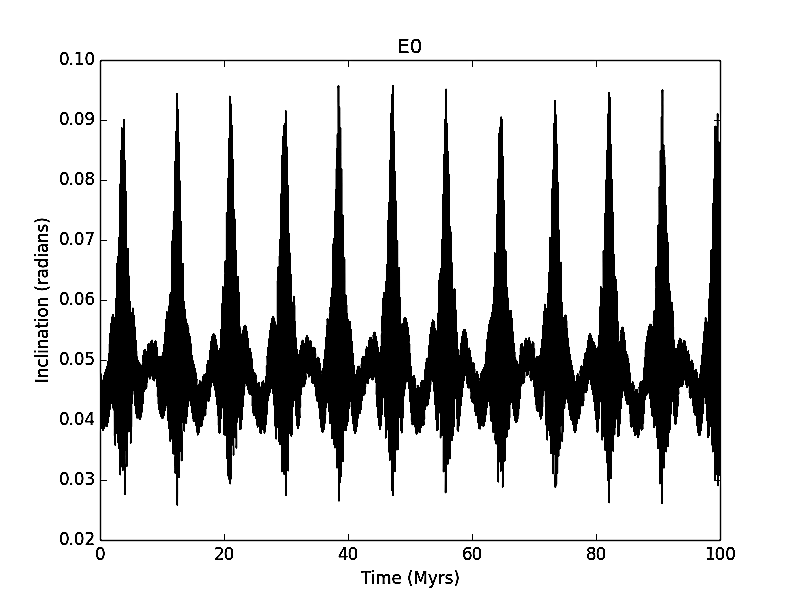}
\qquad
\includegraphics[width=0.5\textwidth]{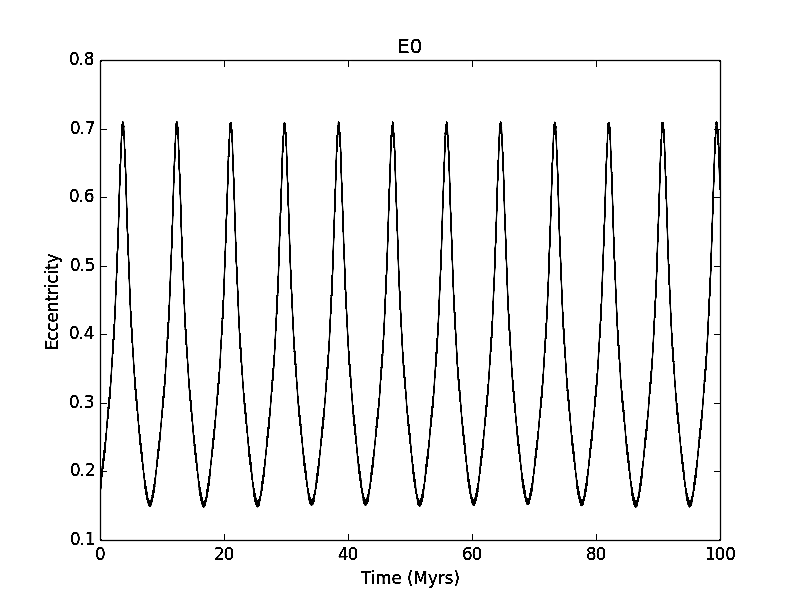}
\qquad
\includegraphics[width=0.5\textwidth]{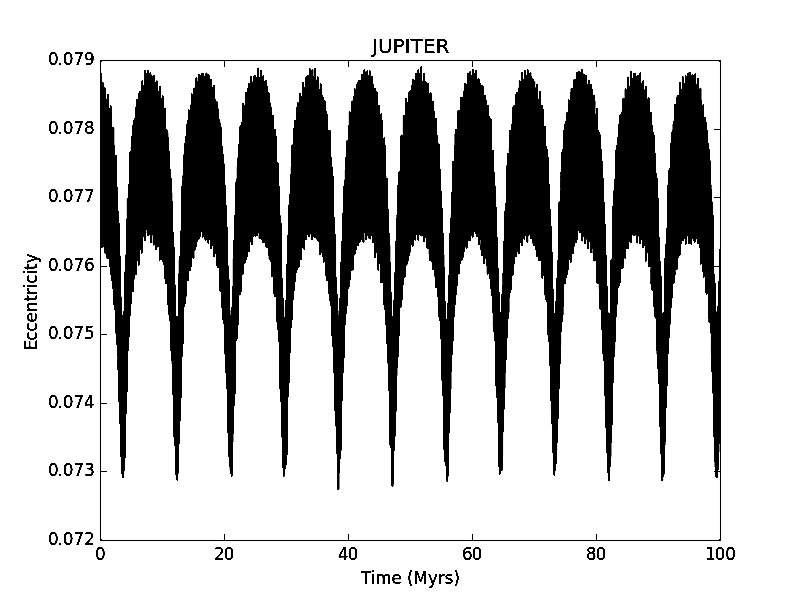}
\caption{Comparison of eejs25's innermost planet's (E0) inclination and eccentricity with Jupiter's eccentricity.  The periods of oscillation match, indicating a strong secularly driven resonance.  Interestingly, this interaction does not drive any chaos in the system.}
\label{fig:eejs25}
\end{figure}

The system with the largest change in AMD over the duration of the simulation is surprisingly non-chaotic.  This outlier (eejs25), is a system of just 2 inner planets.  The innermost planet is $\sim$117\% the mass of Earth, residing at a semi-major axis of 0.62 AU, and the second planet is $\sim$96\% the mass of Venus at a semi-major axis of 1.35 AU.  The innermost planet is locked in a strong, secularly driven resonance with Jupiter (figure \ref{fig:eejs25}).  This causes the eccentricity of the innermost planet to periodically oscillate between $\sim$0.15 and $\sim$0.7 over a period of $\sim$8 Myr.  These oscillations are remarkably stable.  In fact, due to the fortuitous spacing between of the Sun and inner 2 planets, this oscillation does not lead to interactions with other bodies in the system.

\subsubsection{cjs10 and cjs13}
The 2 most stable 5 planet configurations occurred in cjs10 and cjs13, with both systems classified as non-chaotic through runs 1, a and b.  In the runs which vary eccentricity, cjs10 started to develop a weak 5:2 MMR between Jupiter and Saturn, causing mild chaos in run c (where eccentricities are increased by 150\%).  However, the chaos in this run was mild (MEGNO only rose to 3.703 and $\tau_{L}$ for this run was \num{1.57E+09}).  A step increase of 0.05 to Jupiter and Saturn's eccentricity in run e was required to fully introduce chaos (maximum MEGNO values 198.9 and 186.4 for cjs10 and cjs13 respectively) to both of these systems.  This excitation of the eccentricities of the giant planets drove an occasional 3:1 MMR between the second and fourth inner planets in cjs13, possibly contributing to this chaos.  The most remarkable similarity between these systems is their mass spacing and distribution (summarized in Table \ref{table:cjs1013}).  Both have similarly low values of $S_{c}$ (20.1 and 15.6).  In fact, the planet spacing of both systems is somewhat reminiscent of a Titus-Bode Law series.  For example, all orbital locations of cjs13 are within 6\% of a Titus-Bode series beginning at the inner planet's semi-major axis.

\begin{table}
\centering
\begin{tabular}{c c c c c c c}
\hline
     body  & cjs10 a &  cjs13 a & sol a & cjs10 m & cjs13 m & sol m\\ & (AU) & (AU) & (AU) & ($M_{\oplus}$) & ($M_{\oplus}$) & ($M_{\oplus}$) \\
\hline
                1   &   0.51 &    0.54 & 0.39 & 0.61 &  0.68 & 0.055\\
                2   &   0.76 &   0.78  & 0.72 & 0.63 &  0.18 & 0.82\\
                3   &    1.1 & 1.1   &  1.0 &  1.0 &  0.77 & 1.0\\
                4   &    1.9 &      1.6 & 1.5 & 0.46  & 0.59 & 0.11\\
               5   &    2.5 &    2.7 & 2.8 & 0.32  & 0.80 & 0.00015\\
\hline
\end{tabular}
\caption{Comparison of non-chaotic 5 terrestrial planet systems (cjs10 and cjs13) mass and semi-major axis distributions.  The fifth body for the solar system is taken to be Ceres.}
\label{table:cjs1013}
\end{table}

\section{Discussion and Conclusions}

We have presented an analysis of systems of terrestrial planets formed through direct numerical integration of terrestrial accretion, fully evolved to the present epoch.  Our work aims to assess whether our solar system, and it's inherently chaotic dynamics, is a likely result of planetary formation as we currently understand it.  We report that roughly half of our systems display some form of chaos.  By far, the most common source of this dynamical chaos is perturbations from Jupiter.  Additionally, we find that systems in our sample with greater numbers of terrestrial planets are far more prone to chaos than those with fewer inner planets.  Unfortunately, systems formed through numerical integrations (including those of \citet{kaibcowan15} which are used for this work) still routinely produce Mercury and Mars analogues which are far too massive \citep{chambers01, obrien06, chambers07, ray09}.  Consequently, we find it best to consider our solar system a 3 terrestrial planet system for the purposes of comparison in this work.  This classification of the solar system works well with our results that 3-planet systems have an $\sim$50\% chance of being chaotic, and a much lower probability when the giant planets are removed.

By varying the eccentricities of Jupiter and Saturn in 3 separate batches of simulations, we show that an eccentric system of outer planets can quickly introduce dynamical chaos and trigger instabilities in otherwise stable systems.  This result confirms the findings in numerous previous works \citep[e.g.][]{gomes05, mor05, tsi05}.  The inflation in eccentricity required to create such a chaotic system is surprisingly small.  By varying the eccentricity of a batch of systems with Jupiter and Saturn on nearly circular orbits, we show that dynamical chaos quickly ensues.  This sort of event is akin to a Nice Model-like instability \citep{gomes05, mor05, tsi05}.  We go on to show that in such an instability, the possibility of destabilizing the inner planets to the point where a terrestrial planet is lost by either collision or ejection is fairly high.

Additionally, we find that systems most immune to developing dynamical chaos tend to have centers of mass between $\sim$0.8--1.2 AU, though that range is by no means absolute.  This is an interesting result, and likely related to Jupiter's role in driving chaos in many of these systems.

We consistently identified systems throughout our suite of simulations which displayed many of the same chaotic dynamics as our own solar system.  It is clear that chaotic systems such as our own are common results of planetary formation.  The largest source of chaos in our own system, perturbations from Jupiter, is the most common source of chaos observed in our work.  The solar system, however, is akin to only a small fraction ($\sim$10\%) of our simulations since removing just the planets beyond Jupiter turns our system non-chaotic.  Additionally we show that late instabilities are common among these systems, and it is not far-fetched to imagine a late instability shaping dynamics within our own system.  Finally, we find many systems with similar numbers of terrestrial planets, semi-major axis configurations, mass concentrations and chaos indicators ($\tau_{L}$ and MEGNO) as our own.

\section*{Acknowledgements}

This work was supported by NSF award AST-1615975.  Simulations in this paper made use of the REBOUND code which can be downloaded freely at http://github.com/hannorein/rebound.  The bulk of our simulations were performed over a network managed with the HTCondor software package (\url{https://research.cs.wisc.edu/htcondor/}).

\bibliographystyle{apj}
\bibliography{matt1}

\newpage
\appendix
\section{Supplemental Data}

\begin{center}
\begin{longtable}{|c|c|c|c|c|c|c|c|}
\caption[Simulation Results]{Simulation Results} \label{table:data} \\
\hline
\textbf{System} & \textbf{$\tau_{L}$} & \textbf{MEGNO} & \textbf{Inner} & \textbf{Run 1}\footnotemark[1] & \textbf{Run a}\footnotemark[1] & \textbf{Run b}\footnotemark[1] & \textbf{Run 1} \\ & \textbf{(years)} & & \textbf{Planets} & & & & \textbf{MMRs}\footnotemark[2] \\
\hline
\endfirsthead
\multicolumn{4}{c}%
{\tablename\ \thetable\ -- \textit{Continued from previous page}} \\
\hline
\textbf{System} & \textbf{$\tau_{L}$} & \textbf{MEGNO} & \textbf{Inner} & \textbf{Run 1}\footnotemark[1] & \textbf{Run a}\footnotemark[1] & \textbf{Run b}\footnotemark[1] & \textbf{Run 1} \\ & \textbf{(years)} & & \textbf{Planets} & & & & \textbf{MMRs}\footnotemark[2] \\
\hline
\endhead
\hline \multicolumn{4}{r}{\textit{Continued on next page}} \\
\endfoot
\hline
\endlastfoot
      cjs1  & \num{7.09E+05} &     100.7 & 5   &       Y & Y & Y & \\
      cjs2  & \num{1.69E+07} &     45.35 & 4   &       Y & N & N & 7:4\\
      cjs3  & \num{2.93E+05} &     187.2 & 5   &       Y & Y & Y & \\
      cjs4  & \num{3.69E+05} &     160.9 & 5   &       Y & Y & Y & 2:1\\
      cjs5  & \num{4.63E+05} &     183.5 & 5   &       Y & Y & Y & \\
      cjs6  & \num{3.61E+05} &     160.0 & 5   &       Y & Y & Y & \\
      cjs7  & \num{5.04E+06} &     126.0 & 5   &       Y & N & Y & 7:4,5:1,\\
            &                &           &     &         &   &   & 7:1 (S),\\
            &						&	&		&		&	&	&5:2 (J,S)\\
      cjs8  & \num{1.71E+10} &     2.004 & 4   &       N & N & N & \\
      cjs9  & \num{5.93E+10} &     1.997 & 3   &       N & N & N & \\
     cjs10  & \num{8.28E+08} &     1.896 & 5   &       N & N & N & \\
     cjs11  & \num{7.80E+10} &     1.999 & 2   &       N & N & N & \\
     cjs12  & \num{7.14E+06} &     90.97 & 5   &       Y & Y & Y & \\
     cjs13  & \num{6.10E+09} &     2.003 & 5   &       N & N & N & \\
     cjs14  & \num{1.78E+10} &     1.988 & 3   &       N & N & N & \\
     cjs15  & \num{7.61E+07} &     8.794 & 5   &       Y & Y & Y & 7:3 (J)\\
     cjs16  & \num{4.89E+04} &     212.8 & 4   &       Y & Y & Y & \\
     cjs17  & \num{5.97E+08} &     4.467 & 4   &       Y & N & N & 3:1 (J)\\
     cjs18  & \num{5.16E+07} &     21.04 & 3   &       Y & N & N & 3:1 (J)\\
     cjs19  & \num{1.50E+07} &     53.36 & 4   &       Y & Y & Y & \\
     cjs20  & \num{7.23E+10} &     2.000 & 2   &       N & N & N & \\
     cjs21  & \num{9.03E+10} &     1.996 & 2   &       N & N & N & \\
     cjs22  & \num{3.86E+10} &     2.000 & 3   &       N & N & N & \\
     cjs23  & \num{8.83E+03} &     169.0 & 4   &       Y & Y & Y & \\
     cjs24  & \num{1.66E+06} &     20.29 & 4   &       Y & N & Y & 5:2 (J,S)\\
     cjs25  & \num{2.16E+10} &     2.040 & 4   &       N & N & N & 2:1\\
     cjs26  & \num{1.39E+05} &     183.2 & 5   &       Y & Y & Y & 6:1\\
     cjs27  & \num{8.50E+10} &     1.999 & 3   &       N & N & N & \\
     cjs28  & \num{7.22E+10} &     1.999 & 2   &       N & N & N & \\
     cjs29  & \num{9.22E+07} &     12.82 & 4   &       Y & Y & Y & 5:3:1\\
     cjs30  & \num{7.50E+08} &     4.816 & 3   &       Y & N & N & \\
     cjs31  & \num{1.58E+10} &     2.001 & 2   &       N & N & N & \\
     cjs32  & \num{2.81E+09} &     1.966 & 4   &       N & N & N & \\
     cjs33  & \num{1.31E+10} &     1.999 & 2   &       N & N & N & \\
     cjs34  & \num{7.89E+04} &     117.0 & 4   &       Y & Y & Y & \\
     cjs35  & \num{2.26E+10} &     2.000 & 3   &       N & N & N & \\
     cjs36  & \num{3.33E+06} &     186.5 & 4   &       Y & N & Y & \\
     cjs37  & \num{2.14E+05} &     193.5 & 5   &       Y & Y & Y & \\
     cjs38  & \num{3.50E+08} &     5.732 & 3   &       Y & N & Y & \\
     cjs39  & \num{6.30E+10} &     1.999 & 4   &       N & N & N & \\
     cjs40  & \num{6.05E+10} &     1.999 & 4   &       N & N & N & \\
     cjs41  & \num{6.46E+06} &     113.5 & 4   &       Y & Y & Y & 5:3\\
     cjs42  & \num{9.03E+10} &     1.998 & 4   &       N & N & N & 5:1\\
     cjs43  & \num{1.47E+07} &     49.51 & 4   &       Y & N & Y & 4:1\\
     cjs44  & \num{1.93E+05} &     189.4 & 4   &       Y & Y & Y & \\
     cjs45  & \num{1.38E+05} &     194.1 & 4   &       Y & N & Y & \\
     cjs46  & \num{5.10E+10} &     2.011 & 4   &       N & N & N & \\
     cjs47  & \num{2.11E+06} &     129.4 & 4   &       Y & N & Y & \\
     cjs48  & \num{1.68E+08} &     8.483 & 4   &       Y & Y & Y & \\
     cjs49  & \num{1.49E+07} &     20.44 & 4   &       Y & N & Y & \\
     cjs50  & \num{5.08E+08} &     6.271 & 4   &       Y & N & Y & \\
     eejs1    & \num{4.36E+05} &      178.4 & 3         & Y & N & Y & \\
     eejs2   & \num{1.75E+05} &      87.08 & 4         & Y & Y & Y & 5:3\\
     eejs3   & \num{4.55E+05} &      202.9 & 4         & Y & N & Y & 8:5,7:1\\
     eejs4   & \num{9.62E+05} &      174.4 & 1         & Y & N & N & \\
     eejs5   & \num{1.05E+10} &      2.000 & 2         & N & N & N & \\
     eejs6   & \num{4.11E+07} &      29.38 & 3         & Y & Y & Y & 3:1,7:1 (S)\\
     eejs7   & \num{2.41E+07} &      34.65 & 4         & Y & N & Y & \\
     eejs8   & \num{1.33E+10} &      1.988 & 3         & N & N & N & \\
     eejs9   & \num{1.01E+08} &      9.676 & 5         & Y & Y & Y & 8:3,8:5\\     
     eejs10   & \num{6.92E+05} &      151.8 & 1         & Y & N & N & \\
     eejs11   & \num{1.95E+10} &      2.000 & 2         & N & N & N & \\
     eejs13   & \num{9.19E+08} &      7.045 & 4         & Y & Y & N & \\
     eejs15   & \num{2.57E+09} &      1.999 & 1         & N & N & N & \\
     eejs16   & \num{3.39E+10} &      2.002 & 3         & N & N & N & 5:2 (J,S)\\
     eejs18   & \num{1.87E+10} &      2.017 & 4         & N & N & N & \\
     eejs19   & \num{2.93E+10} &      2.000 & 1         & N & N & N & \\
     eejs20   & \num{1.94E+09} &      1.992 & 3         & N & N & N & \\
     eejs21   & \num{4.17E+09} &      1.919 & 3         & N & N & N & \\
     eejs22   & \num{1.05E+10} &      1.998 & 2         & N & N & N & \\
     eejs23   & \num{3.11E+06} &      183.4 & 4         & Y & Y & Y & \\
     eejs24   & \num{1.21E+07} &      57.05 & 5         & Y & Y & Y & 8:1 (S)\\
     eejs25   & \num{6.18E+09} &      2.041 & 2         & N & N & N & \\
     eejs26   & \num{6.31E+09} &      1.912 & 3         & N & N & N & \\
     eejs27   & \num{8.95E+08} &      1.813 & 3         & N & N & N & \\
     eejs28   & \num{2.48E+09} &      4.370 & 3         & Y & N & N & \\
     eejs29   & \num{3.43E+05} &      173.0 & 1         & Y & N & N & 5:2 (J,S)\\
     eejs30   & \num{6.32E+05} &      158.7 & 2         & Y & N & Y & \\
     eejs31   & \num{2.45E+05} &      184.6 & 4         & Y & Y & Y & \\
     eejs32   & \num{1.96E+10} &      1.969 & 3         & N & N & N & 7:3\\
     eejs33   & \num{3.96E+10} &      2.004 & 3         & N & N & N & 7:2,8:5\\
     eejs34   & \num{2.67E+05} &      196.5 & 1         & Y & Y & N & \\
     eejs35   & \num{4.23E+06} &      154.0 & 2         & Y & N & Y & 7:3\\
     eejs36   & \num{3.39E+06} &      180.2 & 1         & Y & N & N & \\
     eejs37   & \num{1.06E+06} &      183.2 & 4         & Y & Y & Y & \\
     eejs38   & \num{2.29E+05} &      171.8 & 2         & Y & N & N & \\
     eejs39   & \num{1.84E+09} &      2.005 & 2         & N & N & N & \\
     eejs40   & \num{1.08E+05} &      189.1 & 4         & Y & Y & Y & \\
     eejs41   & \num{5.27E+06} &      33.10 & 4         & Y & Y & Y & 5:3\\
     eejs42   & \num{1.31E+10} &      2.511 & 4         & N & N & N & \\
     eejs43   & \num{4.94E+09} &      1.998 & 1         & N & N & N & \\
     eejs44   & \num{7.84E+04} &      193.8 & 3         & Y & N & Y & \\
     eejs45   & \num{2.09E+06} &      178.1 & 4         & Y & N & Y & \\
     eejs46   & \num{5.52E+10} &      2.009 & 4         & N & N & N & \\
     eejs47   & \num{5.37E+08} &      8.082 & 3         & Y & N & N & 8:5\\
     eejs49   & \num{3.06E+10} &      1.996 & 3         & N & N & N & 5:2\\
     eejs50   & \num{2.10E+10} &      1.929 & 2         & N & N & N & 7:3\\
ann1 & \num{3.64E+09} & 2.014 & 3 & N & N & Y & \\
ann2 & \num{3.02E+10} & 1.998 & 3 & N & N & N & \\
ann3 & \num{3.06E+10} & 1.988 & 4 & N & N & N & \\
ann4 & \num{4.94E+10} & 2.000 & 3 & N & N & N & \\
ann5 & \num{4.97E+07} & 29.32 & 3 & Y & N & Y & 8:5,7:1 (J) \\
ann6 & \num{1.66E+08} & 7.441 & 4 & Y & N & N & 7:4\\
ann7 & \num{3.91E+09} & 3.132 & 3 & Y & N & N & \\
ann8 & \num{2.87E+05} & 185.6 & 3 & Y & N & Y & \\
ann9 & \num{4.67E+10} & 2.010 & 3 & N & N & N & 2:1\\
ann10 & \num{2.06E+10} & 1.998 & 2 & N & N & N & \\
ann11 & \num{3.17E+09} & 2.294 & 3 & N & N & N & \\
ann12 & \num{6.72E+03} & 195.1 & 4 & Y & Y & Y & \\
ann13 & \num{2.89E+06} & 125.7 & 2 & Y & Y & Y & \\
ann14 & \num{1.68E+07} & 60.09 & 4 & Y & N & N & 4:1\\
ann15 & \num{9.01E+09} & 2.103 & 3 & N & N & N & \\
ann16 & \num{1.85E+10} & 1.99  & 2 & N & N & N & \\
ann17 & \num{1.05E+09} & 4.372 & 4 & Y & N & N & 5:2 (J,S)\\
ann18 & \num{8.23E+08} & 2.159 & 3 & N & N & N & \\
ann19 & \num{3.61E+05} & 183.8 & 3 & Y & Y & Y & \\
ann20 & \num{2.07E+06} & 175.4 & 4 & Y & Y & Y & \\
ann21 & \num{4.55E+06} & 21.76 & 3 & Y & N & Y & 3:1\\
ann22 & \num{2.00E+07} & 44.64 & 3 & Y & N & Y & \\
ann23 & \num{2.00E+07} & 53.55 & 3 & Y & N & Y & 7:4\\
ann24 & \num{2.53E+10} & 1.999 & 3 & N & N & N & \\
ann25 & \num{3.59E+07} & 20.22 & 4 & Y & Y & Y & 8:3\\
ann26 & \num{1.08E+08} & 9.212 & 4 & Y & Y & Y & 5:3\\
ann27 & \num{1.15E+08} & 13.65 & 4 & Y & Y & N & 5:2,8:5\\
ann28 & \num{1.90E+08} & 15.54 & 5 & Y & Y & Y & \\
ann29 & \num{5.97E+06} & 115.7 & 4 & Y & N & Y & \\
ann30 & \num{5.10E+03} & 193.1 & 3 & Y & Y & Y & \\
ann31 & \num{1.32E+08} & 8.47  & 3 & Y & Y & Y & \\
ann32 & \num{1.85E+06} & 182.3 & 4 & Y & Y & Y & \\
ann33 & \num{2.36E+06} & 191.6 & 4 & Y & Y & Y & 2:1\\
ann35 & \num{3.93E+07} & 23.00 & 3 & Y & N & N & \\
ann36 & \num{3.50E+10} & 1.998 & 3 & N & Y & N & 7:3\\
ann37 & \num{2.75E+05} & 192.5 & 4 & Y & Y & Y & \\
ann38 & \num{4.23E+10} & 2.001 & 3 & N & N & N & \\
ann39 & \num{2.50E+05} & 189.6 & 3 & Y & Y & Y & \\
ann40 & \num{6.91E+05} & 192.9 & 4 & Y & Y & Y & \\
ann41 & \num{4.22E+10} & 1.998 & 3 & N & N & N & \\
ann42 & \num{1.01E+10} & 2.000 & 2 & N & N & N & \\
ann43 & \num{1.49E+11} & 2.000 & 3 & N & N & N & \\
ann44 & \num{2.75E+08} & 11.41 & 3 & Y & N & N & \\
ann45 & \num{9.52E+09} & 2.000 & 2 & N & N & N & \\
ann46 & \num{4.81E+06} & 152.4 & 3 & Y & Y & Y & \\
ann47 & \num{3.71E+10} & 2.004 & 3 & N & N & N & \\
ann48 & \num{8.07E+06} & 54.18 & 4 & Y & Y & Y & 5:2 (J,S)\\
ann49 & \num{7.25E+05} & 185.7 & 3 & Y & N & Y & 7:4\\
ann50 & \num{5.53E+10} & 2.008 & 3 & N & N & N & \\
\end{longtable}
\end{center}
\footnotetext[1]{``Y" indicates chaos was detected, ``N" indicates it was not.}
\footnotetext[2]{(J) and (S) indicate the resonance was with Jupiter or Saturn respectively.}

\end{document}